\documentstyle[10pt,aps,prd,amssymb,psfig]{revtex}

\def \be{\begin{equation}}
\def \ee{\end{equation}}
\def \bea{\begin{eqnarray}}
\def \eea{\end{eqnarray}}

\def \s2{\sqrt 2}
\def \sch{Schwarzschild }
\def \ethb{\bar{\eth}}

\begin{document}
\draft

\title{Matter flows around black holes and gravitational radiation}

\author{Philippos Papadopoulos and Jos\'e~A.~Font}

\address{Max-Planck-Institut f\"ur Gravitationsphysik \\
 Albert Einstein Institut \\
Schlaatzweg 1, D-14473, Potsdam, Germany }

\date{\today}

\maketitle

\begin{abstract}
We develop and calibrate a new method for estimating the gravitational
radiation emitted by complex motions of matter sources in the vicinity
of black holes. We compute numerically the linearized curvature
perturbations induced by matter fields evolving in fixed black hole
backgrounds, whose evolution we obtain using the equations of
relativistic hydrodynamics. The current implementation of the proposal
concerns non-rotating holes and axisymmetric hydrodynamical
motions. As first applications we study i) {\em dust shells} falling
onto the black hole {\em isotropically}\, from finite distance, ii)
initially spherical layers of material falling onto a {\em moving}
black hole, and iii) {\em anisotropic} collapse of shells.  We focus on the
dependence of the total gravitational wave energy emission on the flow
parameters, in particular shell thickness, velocity and degree of
anisotropy. The gradual excitation of the black hole quasi-normal mode
frequency by sufficiently compact shells is demonstrated and
discussed. A new prescription for generating physically reasonable
initial data is discussed, along with a range of technical issues
relevant to numerical relativity.
\end{abstract}



\section{Introduction}
\label{intro}

The numerical investigation of relativistic gravity is considered an
important complement to both the analytic studies of strong field
phenomena, and the experimental efforts anticipated to probe such
astrophysical regions. With the exception of the vacuum two-body
problem (i.e., the coalescence of two black holes), all realistic
sources of gravitational radiation involve matter. The joint
integration of the equations of motion for matter and geometry was,
hence, in the minds of theorists from the very beginning of numerical
relativity (e.g., see recollections in~\cite{smarr}). Still, the
effort focused from early-on on vacuum spacetimes, since those clearly
captured, and still do, the least understood part of the
equation. Today the struggle with the perplexities of the ``many
fingered time'' in relativity may be gradually turning in favor of the
numerical relativists, at least for certain classes of spacetimes
(e.g., see recent work in~\cite{BBH1,BBH2,ACS}).

In the meantime, largely motivated by observational data on (special)
relativistic astrophysical phenomena, the numerical investigation of
the relativistic equations of hydrodynamics has made significant
progress. The traditional approach to the numerical integration of the
hydrodynamical equations started with Wilson's pioneering
work~\cite{Wi72}, where the equations were written as a set of
advection equations. Wilson's scheme used a combination of upwind
finite-difference and artificial viscosity techniques to damp spurious
oscillations. Despite its non-conservative character, this approach
has been widely (and successfully) used over the years. Recently,
though, high resolution shock-capturing numerical schemes (HRSC) have
emerged as a solid alternative, which accurately resolves
ultra-relativistic flows. On the theoretical side this progress was
largely based on the skillful exploitation of the hyperbolicity of the
system. On the practical side, the enormous body of work on numerical
non-relativistic hydrodynamics provided a multitude of working
algorithmic techniques, which were successfully transcribed to the
relativistic case (for a review of the current status of the field
see~\cite{betal97,marti-ibanez}). Clearly, handling the matter degrees
of freedom numerically is an increasingly mature subject, which will
in turn help explore large classes of interesting spacetimes.


In this paper we start a line of investigation where we will be
exploring two and three-dimensional relativistic hydrodynamics around
black holes, coupled to {\em approximate} versions of the equations
governing the black hole spacetime geometry. The aim is to produce
simple and robust numerical probes of the gravitational emission of
complex hydrodynamic motions in the vicinity of black holes. Current
and near-future advances in space-borne high-energy instrumentation,
e.g., the Rossi X-Ray Timing Explorer (RXTE) bring the strong field
regions near black holes in increased focus. At the same time, some of
the most ubiquitous sources of gravitational radiation, anticipated to
be detected by the Laser Interferometric Gravitational Observatory
(LIGO), are expected to be binary systems in which a black hole is
either present initially, or forms during the merger. The description
of the violent merger phase is at present understood rather
poorly. The combination of theoretical computations with observational
output will offer, potentially, some thrilling insights into the
relativistic regime. To this end, we propose to use special versions
of black hole perturbation theory, in tandem with full hydrodynamical
simulations, in order to estimate the gravitational wave signal from a
range of systems. The approach attempts (in effect) to eliminate the
need to reproduce and maintain numerically the underlying black hole
spacetime, complementing the on-going efforts for a complete
(non-linear) evolution.

In recent research, {\em vacuum} black hole perturbations were
integrated numerically as 2+1 initial value
problems~\cite{KLP,KLPA}. This work followed the propagation of
generic vacuum initial data through the burst, quasi-normal ringing
and power-law tail phases of the dynamics, confirming anticipated
behavior, even for delicate effects like
super-radiance~\cite{ALP}. The concept of numerical integration of
linearized black hole perturbations is carried with the proposal
presented here one step further, by the inclusion of source terms
generated by extended objects, in particular from fluid material.

Materializing the general proposal, we implement, test and use a
restricted subset, which still covers new ground. More precisely, the
hydrodynamical code is restricted to axisymmetry. The fluid
configurations we study are pressureless gas (dust), albeit evolved
with the full hydrodynamical code. The black hole perturbations
concern a non-rotating hole. We finally focus on the quadrupole part
of the radiation. All of the above restrictions can be lifted with
straightforward extensions of the numerical algorithms.


The organization of the paper is as follows: in Sec.~\ref{physical} we
lay out the physical assumptions that underlie our model.  Next, in
Sec.~\ref{analytic} we compile the mathematical framework that
captures those assumptions. The equations of general relativistic
hydrodynamics, are briefly reviewed here. We describe perturbation
equations with extended sources, in particular hydrodynamical, perfect
fluid, sources. Specializing the discussion, the inhomogeneous
Bardeen-Press equation for perfect fluid matter fields is
analyzed. Two issues pertaining to the selection of initial data are
discussed here, first, an approach to minimizing initial radiation
content, second, practical strategies for avoiding
ambiguities. Section~\ref{numerical} gives a brief description of the
numerical methodology employed for the solution of the hydrodynamical
equations and the inhomogeneous Bardeen-Press equation. The
calibration of the present code is described next. The discussion of
the results (Sec.~\ref{results}) begins with a detailed presentation of a
typical computation. We then present axisymmetric dust computations in
a variety of parametric surveys. The conclusions of this work and a
discussion of technical issues that were raised during this
investigation are to be found in Sec.~\ref{discuss}. The section, and
the paper, close with an outline of future directions.

%
\section{The Physical Model}
\label{physical}
%

The rapid accretion of large lumps of matter onto a black hole emerges
as a fairly generic phenomenon (or at least picture) in
astrophysics. It occurs when the collapse of a massive star gives
birth to a central black hole, as chunks of stellar material are being
swallowed, while others, prevented by angular momentum, are left
orbiting and form a long lived torus. Wholesale accretion may also be
the end act in a black-hole neutron-star merger, where the neutron
star might be swallowed whole, or might first disrupt and then accrete
in pieces. Or, just the same, in a binary neutron star merger, where
the final configuration is too massive to support itself against
collapse and forms a hole surrounded by neutron star matter.

We outline here the idealizations we adopt in order to get a handle on
those intriguing astrophysical events. Those assumptions are fairly
standard in certain fields, but our numerical approach will
be clarified with a recap of the main points.  In our treatment of
accreting systems we take an exact black hole (vacuum) background
spacetime as the stage on which the accretion play unfolds. The
dynamics of an idealized fluid is then explored using the zeroth order
geometry defined by this spacetime. The stress-energy of the moving
fluid generates variable distortions of the spacetime, which
sufficiently far from the black hole (and the matter source) are
interpreted as gravitational waves. The core assumption is then that
the mass $\mu$, of the accreting fluid, is much smaller than the black
hole mass M, i.e., $\mu << M$.

To first order in $\mu$, the matter flow generates metric distortions
which feedback as geometric gravitational forces acting on the
matter. Two of the most important such feedback mechanisms are the
{\em self-gravity} of the fluid, and {\em radiation reaction}
effects. We will neglect in the current programme both mechanisms, by
ignoring the first order metric corrections to the fluid equations of
motion. The first approximation (i.e., no self-gravity) would be valid
for motions at a radius $R$ close to the black hole, where tidal
forces dominate the fluid self-gravity. Assuming a characteristic
fluid size $r$, the radius at which this happens is given
approximately by $R\approx r (M/\mu)^{1/3}$. The second approximation
(i.e., no radiation reaction) would be valid for sufficiently short
intervals of time and weak emission of waves. None of those
approximations is strictly true in most scenarios of astrophysical
interest. Hence we adopt them with the understanding that any new
results pertaining to realistic systems are subject to scrutiny with
respect to their actual applicability.  Still, it has been a
remarkable recent success story, to illustrate the robustness of
black hole perturbation theory as a quantitative analysis
tool~\cite{gleiser1,gleiser2}.

The remaining dominant effects then, governing the fluid motion in our
approximation, are the zeroth-order accelerations and inertial forces
encoded in our choice of a static background spacetime, and,
additionally the {\em pressure gradients} inside the extended
object. Neglecting the pressure gradients is also a common assumption
in the literature, which then leads to the ``cloud of dust particles''
picture and motion along geodesics. Indeed, neither internal stresses
nor inertial forces will compete successfully against gravity in an
{\em advanced stage of accretion}, deep inside the potential well. In
the earlier stages, though, pressure gradients and their interplay
with zeroth-order effects may play an important role. This is
certainly the case for configurations that are in rotating
equilibrium, as for example in the structure of fluid tori orbiting
around black holes. The large number of idealizations we adopted till
this point suggests that a simple description of the fluid
thermodynamics would be sufficient for our purposes. Indeed, the
generally fast timescale of the collapse reduces our concerns about
radiative or dissipative processes. Hence we will work within the
framework of perfect fluids, and for the most part we will be using in
our investigations standard polytropic equations of state, even though
in the applications considered in this paper we actually assume zero
pressure.

We are still left with important sources of complexity. Whereas the
integration of linear equations with potentials can be handled in
certain cases entirely within analytic techniques~\cite{leaver}, the
source terms from realistic extended objects involve steps that appear
to be intractable analytically: i) The time evolution of matter fields
is generally non-linear, with the exception of certain classes of
fundamental fields (e.g., scalar and electromagnetic fields). Simple
initial data for fluids may develop rapidly into solutions with highly
non-linear features (e.g., shock formation in supernova collapse). ii)
The stress-energy tensor is non-linear in the matter fields. iii) The
source term generating curvature waves depends on rather intricate
additions and differentiations of the stress-energy tensor. This
reflects the dependence of gravitational radiation on higher order
moments of the stress-energy. After the $T_{\mu\nu}$ buildup is
complete, those linear operations ``separate the wheat from the
chaff''. 

Radial flows, i.e., flows with purely radial velocity fields, are a
natural subset of the bewildering variety of general flows. Their
relative simplicity will provide us here with a testing ground for our
algorithms and a starting point towards more general cases. In the
case of dust matter, we may further decompose radial motions into {\em
isotropic radial flows}, where the pattern of fluid velocity does not
have any angular dependence (hence all emission comes from the
accelerated motion of some angularly structured density), and {\em
anisotropic radial flows} in which the velocity field is direction
dependent.  Finite-sized collections of dust, shaped in the form of
stars or shells, falling onto the black hole isotropically, have been
studied quite
extensively~\cite{sana81,hsw82,shawa82,oona83a,oona83b,peshawa85}.
These studies brought to the foreground the fact that for a fixed
amount of infalling mass the gravitational radiation efficiency is
reduced compared to the point particle limit. In fact, those studies
indicate that the energy emitted by infalling dust will never exceed
that of a particle with the same mass \cite{shawa82,peshawa85}. The
reduction is, intuitively, due to cancellations of the emission from
distinct parts of the extended object. It appears that the study of
increasingly realistic {\em extended} objects is crucial for an
accurate assessment of the waveform of the generated waves in the late
stages of a binary merger.

Within the assumptions mentioned, and the methodological restriction
to radial dust flows, our selection of initial conditions attempts to
cover the range of qualitatively different possibilities.  We focus on
highly resolved radial structures and attempt to keep the angular
dependence to the minimal necessary.  We anticipate that this trend
should be reversed in configurations with angular momentum.

%
\section{Analytic Framework}
\label{analytic}
%

\subsection{Perfect fluid hydrodynamics in background black hole spacetimes}

The motion of matter fields in a general curved spacetime is governed
by the local conservation laws of baryon number and energy-momentum
\begin{equation}
\nabla_{\mu}J^{\mu}=0, \,\,\,\,\,\, \nabla_{\mu}T^{\mu\nu}=0,
\end{equation}
where $J^{\mu}=\rho u^{\mu}$ is the density current, $T^{\mu\nu}=
\rho h u^{\mu}u^{\nu} + p g^{\mu \nu}$ is the stress-energy tensor for
a perfect fluid and $\nabla_{\mu}$ is the covariant derivative associated
with the background metric $g_{\mu\nu}$. 
In those expressions all quantities have their usual meaning, that is,
$\rho$ is the rest-mass fluid density, $h$ is the specific enthalpy, defined as
$h=1+\epsilon+ p/\rho$, $\epsilon$ is the specific internal energy,
$p$ is the pressure and $u^{\mu}$ is the fluid 4-velocity.  The system of
equations must be closed with an appropriate equation of state,
$p=p(\rho,\epsilon)$. Greek (Latin) indices run from 0 to 3 (1 to 3)
and we use natural units $(G=c=1)$ throughout.

By choosing an appropriate set of variables, the equations of general
relativistic hydrodynamics are written as a hyperbolic system of
balance laws. In~\cite{betal97} this is done by defining quantities
which are directly measured by Eulerian observers, i.e., the rest-mass
density $D = \rho W$ , the momentum density in the $j$-direction $S_j
= \rho h W^2 v_j $ and the total energy density $E = \rho h W^2 - p$.
In these expressions $W$ stands for the Lorentz factor, which
satisfies $W=(1-{v}^{2})^{-1/2}$ with ${v}^{2}= \gamma_{ij} v^i v^j$,
where $v^i$ is the 3-velocity of the fluid, defined as $v^i= u^i/W +
\beta^i/\alpha$, where $\alpha$ and $\beta^i$ are the spacetime lapse
function and shift vector, respectively, and $\gamma_{ij}$ are the
spatial components of the spacetime metric where the fluid evolves:
\begin{equation}
ds^{2} = -(\alpha^{2}-\beta_{i}\beta^{i}) dt^{2}+
2 \beta_{i} dx^{i} dt + \gamma_{ij} dx^{i}dx^{j} \, .
\end{equation}
\noindent
For a generic spacetime the system of equations we solve reads
\begin{equation}
\frac{1}{\sqrt{-g}} \left(
\frac {\partial \sqrt{\gamma}{\bf U}({\bf w})}
{\partial t} +
\frac {\partial \sqrt{-g}{\bf F}^{i}({\bf w})}
{\partial x^{i}} \right)
 = {\bf S}({\bf w}) \, ,
\label{F}
\end{equation}
\noindent
where $g\equiv \det(g_{\mu\nu})$ is such that
\begin{displaymath}
\sqrt{-g} = \alpha\sqrt{\gamma} ,\,\,\,\,\,\,\,\,\,
\gamma\equiv \det(\gamma_{ij}) \, ,
\end{displaymath}
\noindent
and {\em det} stands for the determinant of the corresponding matrix. 
In this work we use the standard form of the black hole metric in
Boyer-Lindquist coordinates $(t,r,\theta,\phi$), so the above expressions
are specialized accordingly.

In Eq.~(\ref{F}) the state vector of {\it primitive} variables is defined
as ${\bf w} = (\rho, v_{i}, \varepsilon)$ and the vector of {\it
evolved} variables is ${\bf U}({\bf w}) = (D, S_j, \tau)$, with $\tau
\equiv E - D$, that is the total energy density subtracting the
rest-mass density.  In addition, the {\it fluxes} are
\begin{equation}
{\bf F}^{i}({\bf w})  =   \left(D \left(v^{i}-\frac{\beta^i}{\alpha}\right),
 S_j \left(v^{i}-\frac{\beta^i}{\alpha}\right) + p \delta^i_j,
\tau \left(v^{i}-\frac{\beta^i}{\alpha}\right)+ p v^{i} \right) \, ,
\end{equation}
\noindent
and, finally,
the corresponding source terms ${\bf S}({\bf w})$ are given by
\begin{eqnarray}
{\bf S}({\bf w}) =  \left(0,
T^{\mu \nu} \left(
\frac {\partial g_{\nu j}}{\partial x^{\mu}} -
\Gamma^{\delta}_{\nu \mu} g_{\delta j} \right),
\alpha  \left(T^{\mu t} \frac {\partial {\rm ln} \alpha}{\partial x^{\mu}} -
T^{\mu \nu} \Gamma^t_{\nu \mu} \right)
                     \right),
\end{eqnarray}
\noindent
where $\Gamma^{\delta}_{\nu \mu}$ are the 4-dimensional Christoffel
symbols. 

The physical boundary conditions we impose on the hydrodynamic
quantities depend largely on the scenario under study. In all cases
considered in this work we adopt ingoing radial boundary conditions
(towards the hole), both at the horizon and at the outermost zone.
The former condition (at the horizon) is strictly valid as the fluid
accretes onto the black hole supersonically. The latter condition
(outer boundary) implies a continuous inflow of very {\em low density}
matter through the outer boundary of the domain. In the angular
direction, axisymmetry considerations prescribe the appropriate
conditions at the poles ($\theta=0$ and $\pi$).

\subsection{Curvature perturbations by hydrodynamical sources}

A direct integration of a perturbative system of the Einstein
equations has to follow closely, methodologically, the corresponding
non-linear approaches, and actually carries a similar computational
load, along with gauge and boundary problems.  We choose instead to
base our approximation of the black hole geometry on the highly
developed theory of curvature perturbations (the subject is
comprehensively exposed in the monumental book of
Chandrasekhar~\cite{chandra}). This formalism is teeming with positive
features: the evolved quantities are invariant under infinitesimal
coordinate and tetrad transformations~\cite{stewart}, all possible
types of perturbations are neatly captured in the real and imaginary
parts of the Weyl tensor scalar, and rotating black holes are covered
by simple extensions of the non-rotating case. Last, but not least,
the formalism leads quite naturally to partial differential equations
defined in the time domain.

The dynamics of the Weyl spinor $\Psi_{ABCD}$ is given
generally~\cite{penrose} by
\begin{equation}
\nabla^{A}_{B'} \Psi_{ABCD} = 4 \pi \nabla^{A'}_{(B} T_{CD)A'B'} \, , 
\end{equation}
which shows that spacetime derivatives of the stress-energy tensor
$T_{\mu\nu}$ act as sources to the tidal field $\Psi$. Those evolution
equations for the $\Psi$ components are valid for any spacetime
satisfying the Einstein equations, and reflecting their origin in the
Bianchi identities, are ``already linearized'' in the dependent
quantities.  Their structure is similar to the propagation equations
for fields of other spin, most notably the Maxwell field. The
important difference is that the metric, implicit in the connection
$\nabla^{A}_{B'}$, is here part of the theory and must be propagated
forward, along with a host of other variables in order to close the
system of equations.

The concept of evolving the curvature components becomes radically
simpler in the framework of the Newman-Penrose formalism, when applied
to black hole perturbations~\cite{chandra,stewart}. For a general
class of exact solutions, the equations for the {\em perturbed} Weyl
tensor components ($\delta\Psi$) decouple, and when written in the
frequency domain even separate~\cite{teuk72,teuk73}. Here we will use
the limiting case for non-rotating black holes, i.e., the
inhomogeneous Bardeen-Press (BP) equation~\cite{bardeen}.  We adopt
the standard form of the equation, i.e., Boyer-Lindquist coordinates
and the Kinnersley null tetrad $(l^{\mu},n^{\mu},
m^{\mu},\bar{m}^{\mu})$. We consider solely the spin $-2$ field, which
encodes, at large $r$, the outgoing radiation. For brevity we use the
operators
\begin{eqnarray}
L_{+} & = & \frac{\partial}{\partial t } + 
            \frac{\partial}{\partial r_{*}}  \, ,  \\ 
L_{-} & = & \frac{\partial}{\partial t } - 
            \frac{\partial}{\partial r_{*}}  \, , \\ 
\eth  & = & - \frac{\partial}{\partial \theta} + s \cot{\theta} 
            - \frac{i}{\sin\theta}\frac{\partial}{\partial \phi}  \, ,\\ 
\ethb & = & - \frac{\partial}{\partial \theta} - s \cot{\theta}
            + \frac{i}{\sin\theta}\frac{\partial}{\partial \phi} \, ,
\end{eqnarray}
where $\eth$,$\ethb$ are the standard eth and eth-bar operators (see
e.g.,~\cite{penrose,goldberg}) and $s$ is the {\em spin-weight} of the
field acted upon by the angular operators. We use the tortoise
coordinate $r_{*}$, defined by $dr_{*} = (r^2/\Delta) dr$, where
$\Delta$ is the {\em horizon function} $\Delta=r^2-2Mr$ with $M$ being
the mass of the black hole.  We use as dependent variable the complex
field $Y$, related to the Weyl tensor tetrad component
$\Psi_4=-C_{\mu\nu\rho\sigma}
n^{\mu}\bar{m}^{\nu}n^{\rho}\bar{m}^{\sigma}$ by $Y=r\Psi_4$. With
those choices, the BP equation reads
\begin{equation}
\Box Y = 8 \pi \frac{\Delta}{r} T_{4},
\label{BP}
\end{equation}
where the ``wave'' operator expands as
\begin{equation}
\Box = L_{-}L_{+} 
- \frac{4(r-3M)}{r^2} L_{+} 
+ \frac{6M\Delta}{r^5} 
- \frac{\Delta}{r^4} \ethb \eth . 
\label{BP-box}
\end{equation}

The source term of equation~(\ref{BP}), transcribed from~\cite{teuk73} in
our notation, is
\begin{eqnarray}
T_{4} & = & - \frac{1}{2r^2} \ethb \ethb T_{nn} 
        - \frac{1}{4} L_{-}^{2} T_{\bar{m}\bar{m}}
        - \frac{\Delta^2}{r^6} T_{\bar{m}\bar{m}} \\ \nonumber
& &
        + \frac{3r-5M}{2r^2} L_{-} T_{\bar{m}\bar{m}}
        + \frac{3r-8M}{2\sqrt{2}r^3} \ethb T_{n\bar{m}}
        - \frac{1}{\sqrt{2}r} L_{-} \ethb T_{n\bar{m}} \, , 
\label{source}
\end{eqnarray}
where $T_{\hat{a}\hat{b}}$ are the contractions of the stress-energy
tensor with the null-tetrad.

The analytical problem should be clear at this point: given initial
data $Y,\dot{Y}$ on an initial spacelike hypersurface $t_{0}$, and a
spacetime function $T_{4}(x^{\mu})$, compute the value of $Y$ to the
future of $t_{0}$. We further assume that $T_{4}$ has compact support
and we are interested, in particular, in the ``time signal"
$Y(t,r_{o})$, where $r_{o}$ is some remote observer location. The
limit case where the source is non-zero only along a timelike
trajectory has been studied extensively with semi-analytic methods,
mostly in connection with estimates of inspiraling compact binary
waveforms (see e.g.,~\cite{poisson,tanaka}).

The higher derivatives of metric quantities involved in the curvature
approach to perturbations limit the amount of geometric information
that can be readily extracted in time-domain integrations of the
decoupled equations. Assuming the components of the Weyl curvature
tensor have been reconstructed, the complete {\em metric information}
is in principle extractable~\cite{chandra} but in practice may be
cumbersome to compute, especially in the rotating black hole
case. This is, of course, not a problem at infinity, where the
gravitational wave signal is to be extracted.

The asymptotic behavior of the field $Y$ along ingoing and outgoing
null directions is important for numerical purposes, as it reflects
the radial behavior of generic solutions starting with data of compact
support. The asymptotics of outward propagating solutions to the {\em
homogeneous} equation are given by
\begin{equation}
\lim_{r^* \to +\infty}{Y} \sim  1  , \,\,\,\,
\lim_{r^* \to -\infty}{Y} \sim  1  \, , 
\label{asymp-out}
\end{equation}
while inward propagating solutions behave as
\begin{equation}
\lim_{r^* \to +\infty}{Y}  \sim  r^2 , \,\,\,\,
\lim_{r^* \to -\infty}{Y}  \sim  \Delta^{2} \; .
\label{asymp}
\end{equation}

The exponential radial decay (in $r_{*}$) of inward propagating
solutions near the horizon implies $Y = 0$ there.  In practice,
assuming pure ingoing waves, ($L_{-}Y=0$), at a finite but very small
distance from the horizon works well.  The asymptotic dependence is
not as favorable in the outer boundary (this issue is discussed
further in Sec.~\ref{discuss}).  In practice we consider {\em only}
the domain of dependence of the initial data.  In two-dimensional
simulations the symmetry axis is handled with simple regularity
conditions.

The analytic formulation of the evolution equation for the $s=2$
component of the Weyl tensor is entirely analogous to the $s=-2$ case
and it would appear that similar algorithms would be effective.
Nevertheless, the limiting behavior of this fields near the horizon
leads to numerical problems~\cite{KLPA} (already present in radial
integrations in the frequency domain~\cite{teuk-press}).

The contractions of the perfect fluid stress-energy tensor with the
null tetrad read $T_{nn} = \rho h u_{n}^2$, $ T_{n\bar{m}} = \rho h
u_{n} \bar{u}$ and $ T_{\bar{m}\bar{m}} = \rho h \bar{u}^2$, where we
introduced the spin-weighted functions $u_{n}=u^{\mu}n_{\mu}, u=
u^{\mu}m_{\mu}$ and $\bar{u}=u^{\mu} \bar{m}_{\mu} $ with spin-weight
$0,1,-1$ respectively. Some interesting simplifications that have
emerged from the use of the Newman-Penrose formalism are notable
here. The entire physical content of the flow, to the extend that it
affects $\Psi_4$, is captured by the four scalar functions
$(\hat{\rho},u_{n},u)$ ($u$ is complex), where $\hat{\rho}=\rho
h$. 

We single out the sub-case $u_{n} \neq 0, u = 0$, which corresponds to
{\em purely radial motion}. For such motions only the $T_{nn}$ term in
Eq.~(\ref{source}) remains.  It is worth noting that the other extreme
geometry, of {\em purely rotational motion}, as is for example
approximately the case in a neutron-star black-hole binary, does not
zero-out the $T_{nn}$ term, as $T_{nn}$ is bounded below by the rest
mass density. In such rapidly rotating material the emission will be
governed by the combined effect of the time variability in the angular
structure of $T_{nn}$ {\em and} the strong fluid currents encoded in
$T_{\bar{m}\bar{m}}$.

In the case of purely radial motion, we have the further subdivision
according to whether the radial velocity
($v=\sqrt{\gamma_{ij}v^{i}v^{j}}$) is isotropic or not, i.e., whether
$\eth v = 0$. In the general (anisotropic) case the radial motion
source term expands as
\begin{equation}
\ethb\ethb T_{nn} = \ethb\ethb\hat{\rho} \, u_{n}^2
+ 4 u_{n} \ethb\hat{\rho} \, \ethb u_{n} 
+ 2 u_{n} \hat{\rho} \,  \ethb \ethb u_{n} + 
2 \hat{\rho} \, \ethb u_{n} \ethb u_{n}.
\label{tnn}
\end{equation}
The contributions to the source term can now be analyzed with respect
to their angular geometry. In the isotropic case all wave emission is
induced by the angular profile of the density
($\ethb\ethb\hat{\rho}$), although its intensity will also depend on
the magnitude of the radial velocity (the $u_{n}^2$ factor).  It is
easily seen that in this case, the source separates completely, and
different multipoles of the density directly excite the different
multipoles of radiation. This can be shown explicitly by e.g., raising
the spin of the equation twice and then expanding the function $R =
\eth\eth Y$ in spherical harmonics to reach the form
\begin{equation}
\Box_{l} R_{lm} = 8 \pi \frac{\Delta}{r} u_{n}^2 
\left[l(l-1)(l+1)(l+2)\right] \hat{\rho}_{lm} \, ,
\label{despun1}
\end{equation}
where
\begin{equation}
\Box_{l} = L_{-}L_{+} 
- \frac{4(r-3M)}{r^2} L_{+} 
+ \frac{6M\Delta}{r^5} 
- \frac{\Delta}{r^4} l (l+1) \, .
\label{despun2}
\end{equation}

Assuming angular dependence for the radial velocity generates angular
dependence for $u_{n}$ and hence activates all terms in
Eq.~(\ref{tnn}). For example, a spherically symmetric shell would
still generate a non-zero source term if the radial velocity of the
fluid had at least a dipole component. Expanding the expression of the
fourth term in Eq.~(\ref{tnn}) in spin-weighted harmonics shows that
this combination produces a spin (-2) quadrupole. Similarly, a dipole
density distribution is seen to couple with a dipole moment in the
velocity to produce an additional lowest order contribution to the
quadrupole emission. This interesting non-linear complication,
introduced by anisotropy, does not appear to have been analyzed so far
in studies of matter collapse onto black holes. The effect of such
leading order anisotropies (i.e., dipole) will be probed in
section~\ref{results}. One final comment on the structure of the
source term for fluids, in particular radial flows, concerns the types
of perturbations excited by the sources. We see that radial motions in
axisymmetry generate only real perturbations, the imaginary part of
the field being associated with non-axisymmetric patterns of
density/radial velocity.

\subsection{Initial data for the perturbation fields}

We outline here our procedure for selecting initial data for the
perturbation fields. The issue can be of outmost relevance for
numerical investigations, as spurious behavior may completely mask the
sought-after signal. It will be demonstrated in Sec.~\ref{results}
that our concrete computations are performed in a way that bypasses
the issue. Here we summarize and discuss the evidence we accumulated
in the process of dealing with this issue.

Although $\Psi_4$ is typically associated with propagating
gravitational radiation, it is generally non-zero and non-propagating
in regions occupied by matter. For example, in equilibrium matter
configurations around black holes, which are possible with support
from angular momentum, say in the form of tori, the BP equation admits
also {\em static} solutions. Physically, those solutions represent the
first order deviation of the curvature of the equilibrium system from
that of the underlying vacuum spacetime. Clearly, in this case, the
initial data selection procedure should accommodate for this static,
``Coulombic'', part of $\Psi_4$. In this paper we focus exclusively on
purely radial motions with zero angular momentum. Here, there can be
no true static solutions. Still, the initial distribution of density
should be complemented physically with a non-zero curvature, which,
without being static, should only change appreciably in timescales
associated with those of the flow (which we can arrange initially to
be very slow). Prescribing values for the initial curvature
$Y,\dot{Y}$, once an initial distribution of $T_{4}$ is given, such
that there is no ``unseemly" initial burst of radiation, constitutes
the problem of specifying ``physical'' initial data.

We investigate the possibility of utilizing the concept of an
``approximately'' static curvature for setting initial data. We
computed such solutions by considering the {\em elliptic} problem
obtained from the BP equation, upon assuming static curvature fields.
In regular integrations we achieved an entirely equivalent effect by
considering the field $Z=\dot{Y}$ and the corresponding equation
\begin{equation}
\Box Z = 8 \pi \frac{\Delta}{r} \dot{T}_{4},
\label{BP_dot}
\end{equation}
which is straightforward to obtain from the time derivative of
Eq.~(\ref{BP}) above, since all operators commute with $\partial_{t}$.
In all the runs shown below, the initial wave data are assumed to be
zero throughout the domain, i.e., $Z=0,\dot{Z}=0$.  When using such
data, the development of non-zero values for the time derivative $Z$
of the tidal field $Y$ is now associated with the time derivative of
the source term $T_4$. Since those time derivatives can be made zero
initially, one would expect intuitively that this prescription would
better control the response of the curvature to the slowly
accelerating motion of the matter. We find indeed that this approach
significantly reduces the unphysical initial radiation content of the
data {\em compared} to setting zero data on $Y$.

The radial infall scenarios are generally weak emitters, and hence in
the configurations investigated in this work the likelihood of
contamination is high. Indeed we find that even with the significant
reduction achieved with the above recipe, several configurations
produce real signals with amplitude below that of the initial
burst. In the computations below, we opt to {\em leave in the past}
any ambiguities associated with the choice of initial data. We run
configurations which, by construction, do not emit appreciably except
at late times, hence isolating the two problems. A recent
study~\cite{gleiser3} illustrates another clear case of significant
radiation hidden in the choice of initial data.

%
\section{Numerical Method}
\label{numerical}
%

\subsection{Integration of the hydrodynamic equations}

We solve the equations of relativistic hydrodynamics on a discrete
numerical grid with a state-of-the-art HRSC scheme. The procedure has
been described in~\cite{betal97} and successfully applied
in~\cite{font1,font2}. We briefly summarize pertinent details in this
section, referring the interested reader to~\cite{betal97} for further
references, details on code calibration, numerical tests and
algorithmic issues.

Assuming axisymmetry (2D) and Boyer-Lindquist coordinates
$(t,r,\theta,\phi)$, the vector of evolved quantities, is updated from
time level $t^n$ to $t^{n+1}$ according to the following conservative
algorithm
\begin{eqnarray}
{\bf U}^{n+1}_{i,j} = {\bf U}^{n}_{i,j} - \frac{\Delta t}{\Delta r}
  \left[ \hat{\bf F}^{r}_{i+\frac{1}{2},j} - \hat{\bf
  F}^{r}_{i-\frac{1}{2},j} \right] - \frac{\Delta t}{\Delta\theta}
  \left[ \hat{\bf F}^{\theta}_{i,j+\frac{1}{2}} - \hat{\bf
  F}^{\theta}_{i,j-\frac{1}{2}} \right] + \Delta t {\bf S}_{i,j} \, ,
\label{discre}
\end{eqnarray}
\noindent
where $\Delta t = t^{n+1} - t^n$, and $\Delta r$ and $\Delta\theta$
indicate the radial and angular grid spacing, respectively.  In
addition, indices $i$ and $j$ label the radial and angular zones,
respectively.  In Eq.~(\ref{discre}), ${\bf U}_{i,j}$ and ${\bf
S}_{i,j}$ are the mean values of the state and source vector in the
corresponding two-dimensional cell, while $\hat{\bf
F}^{r}_{i+\frac{1}{2},j}$ and $\hat{\bf F}^{\theta}_{i,j+\frac{1}{2}}$
are the {\it numerical fluxes} which are computed at the
cell interfaces.  These fluxes are calculated
using an {\it approximate} (linearized) Riemann solver built upon the
characteristic information of the system~\cite{betal97}.
The canonical flux-formula we use reads
\begin{eqnarray}
\hat{\bf F}_{i\pm 1/2} = \frac{1}{2}
\left({\bf F}_{i\pm 1/2}({\bf w}_{\rm R}) +
 {\bf F}_{i\pm 1/2}({\bf w}_{\rm L}) -
 \sum_{n=1}^4 |\widetilde{\lambda}_{(n)} |
 \Delta \widetilde{\omega}_{(n)}
\widetilde{\bf r}_{(n),{i\pm 1/2}} \right) \, , 
\label{nflux}
\end{eqnarray}
\noindent
where ${\bf w}_{\rm L}$ and ${\bf w}_{\rm R}$ are the values of the
primitive variables at the left and right sides of a given cell
interface. They are derived by a monotonic linear reconstruction of
their cell-centered values~\cite{vl79}. This procedure ensures, in
absence of shocks, second-order accuracy in space.  In
Eq.~(\ref{nflux}), $\{\widetilde{\lambda}_n, \widetilde{\bf
r}_n\}_{n=1,..,4}$ are the eigenvalues and right-eigenvectors of the
Jacobian matrices of system~(\ref{F}), evaluated at the cell-interface
using the arithmetic mean of the left and right states.  The jumps of
the characteristic variables across each characteristic field,
$\{\Delta \widetilde{\omega}_n\}_{n=1,..,4}$, are given by
\begin{equation}
{\bf U}({\bf w}_{\rm R})-{\bf U}({\bf w}_{\rm L}) =
\sum_{n=1}^4 \Delta \widetilde{\omega}_n \widetilde{\bf r}_n.
\end{equation}
\noindent
The code makes use of a third-order TVD (total-variation-diminishing)
Runge-Kutta scheme to perform the time integration algorithm of
Eq.~(\ref{discre}), which is done simultaneously in both spatial
directions. Finally, a one-dimensional Newton-Raphson iteration leads,
at each time step, from the evolved variables to the primitive
ones~\cite{MM2}.

The well known coordinate singularities of the Boyer-Lindquist
coordinates introduce some awkwardness in numerical work. Pure inflow
conditions for the fluid cannot be applied arbitrarily close to the
horizon, as the coordinate velocities become singular there. In the
scenarios considered here, the supersonic nature of the flow ensures
that simple inflow conditions would work well. We have tested that
posing the boundary conditions closer to the horizon (a range from
$r_{*}=-2$ to $r_{*}=-3$) does not affect the flow pattern and the
corresponding waveforms. This issue is discussed further in
Sec.~\ref{discuss}.

\subsection{A simple differencing scheme for the Bardeen-Press equation}

Successful finite difference discretizations of the equation in the
homogeneous case have been published in~\cite{KLPA}. We present here a
somewhat simpler algorithm which we implemented and tested in two
different guises: i) A 2+1 finite difference computation in which
angular derivatives are computed numerically, ii) A multipole
decomposition approach leading to systems of (decoupled in the
non-rotating case) 1+1 equations. Whereas the first approach easily
generalizes to rotating holes and higher dimensions, efficiency
concerns led us to focus here on the second approach, i.e., separating
into multipoles, and considering only the low-lying ones.  The natural
extension of the multipole decomposition to the rotating black hole
case is through a pseudo-spectral algorithm for the angular terms.

The BP equation, Eq.~(\ref{BP}), is discretized on a uniform tortoise
radial grid using a straightforward implementation of the three-level
leapfrog method. The radial and angular derivatives are approximated
as
\begin{eqnarray}
\label{bp-rdiffs}
 \left[Y_{,\theta\theta}\right]^{n}_{i,j} & = & 
 ( Y^{n}_{i,j+1} - 2 Y^{n}_{i,j}  + Y^{n}_{i,j-1})/ (\Delta\theta)^2 \,,
\nonumber \\
 \left[Y_{,\theta}\right]^{n}_{i,j} & = & 
 ( Y^{n}_{i,j+1}  -  Y^{n}_{i,j-1} ))/ (2 \Delta\theta) \,, \nonumber  \\
  \left[Y_{,r_{*}}\right]^{n}_{i,j}  & = & 
 ( Y^{n}_{i+1,j} - Y^{n}_{i-1,j}) / (2 \Delta r_{*}) \,, \nonumber \\
  \left[Y_{,r_{*}r_{*}}\right]^{n}_{i,j}  & = & 
 ( Y^{n}_{i+1,j} - 2 Y^{n}_{i,j} +   Y^{n}_{i-1,j} )/ (\Delta r_{*})^2.
\end{eqnarray}
The time derivatives are approximated as
\begin{eqnarray}
\label{bp-tdiffs}
 \left[Y_{,tt}\right]^{n}_{i,j} & = & 
 ( Y^{n+1}_{i,j} - 2 Y^{n}_{i,j}  + Y^{n-1}_{i,j})/ (\Delta t)^2 \,, \nonumber \\
 \left[Y_{,t}\right]^{n}_{i,j} & = & 
 ( Y^{n+1}_{i,j}  -  Y^{n-1}_{i,j} ))/ (2 \Delta t).
\end{eqnarray}
Substituting those approximated expressions in the BP equation and
solving for the discrete variable $Y^{n+1}_{i,j}$ gives an explicit
update routine, which is stable subject to the usual CFL condition.

A direct implementation of the
scheme~(\ref{bp-rdiffs},\ref{bp-tdiffs}) leads at late times (of the
order of hundreds of $M$) to a high frequency instability which is
seen to emerge from the area around $r=3M$. Upon performing a frozen
coefficient stability analysis of the BP equation, in this region, it
can be seen that {\em local} propagating modes are unstable even in
the continuum limit. This instability is inherited to the discrete
equations, where, in contrast to the analytic canceling of the blowup
as local modes propagate away from the region, the numerical modes
grow uncontrollably. The localized and high frequency nature of the
instability leads to simple cures, e.g., by using a numerical scheme
with higher intrinsic viscosity, or by explicitly adding some
artificial viscosity to the discretized equation (the option we followed
here).

The inner boundary conditions are those of pure ingoing waves.  If
those conditions are applied sufficiently close to the horizon, this
approximation of Eq.~(\ref{asymp}) is very accurate. In terms of the
Boyer-Lindquist $r$ coordinate our typical location of the inner
boundary corresponds to a coordinate distance of $|r-r_+| \approx
10^{-14}$, leading to $\Delta^2 \approx 10^{-28}$.

For reasons of numerical efficiency, we have also developed a finite
difference implementation of the BP equation which explicitly
separates $Y$ into spin harmonics. In the non-rotating black hole case
each multipole evolves independently, under the influence of the
corresponding multipole source term, which we construct numerically from
the function $T_{4}$ by performing a projection onto the corresponding
spin harmonic:
\begin{equation}
T_{lm}(t,r_{*}) = \int d\Omega \, {}_{-2}\bar{Y}_{lm} 
T_{4}(t,r_{*},\theta,\phi).
\end{equation}

Reflecting our physical assumptions and the mathematical model, the
coupling of the numerical algorithms is {\em one-directional}, i.e.,
numerical information generated by the hydrodynamical code is
converted into appropriate numerical sources to the BP equation.  The
generated waveform does not feedback into the fluid motion.  An
important feature of the numerical coupling is that the numerical
evolution of the two systems is not done on grids of the same radial
extend. As already mentioned, the boundary conditions for the wave
equation are those of pure ingoing waves, and can be applied (and work
well) sufficiently inside the potential well. A value of $r_{*}=-10$
is more than sufficient and we typically use $r_{*}=-50$. In contrast,
the inner boundary for the fluid must be kept at somewhat larger
radius. The situation is reversed in the outer boundary: Appropriate
treatment of the asymptotic region is much easier for the hydrodynamic
quantities than for the conformal field $Y$. This leads to the
adoption of a grid considerably more extended in radius for the wave
compared to the hydro grid which is kept to the minimal sufficient
range. Finally, the computation of gravitational wave energy $E$ is
performed with the integration of the outgoing signal in time at a
fixed observer radius, using a second order accurate summation over
time and angles, and we use the {\em baryonic} mass, integrated over
the initial hypersurface, as an estimate for the total mass $\mu$.

\subsection{Calibration of the coupled algorithm}

We demonstrate here convergence results of the gravitational waveform
for increasingly finer grid resolution. Although a formal convergence
analysis can be done using any finite integration time, even if small,
we choose instead to check convergence for our typical total evolution
times, i.e., of the order of half thousand $M$. This raises the
computational requirements for performing the test, but we considered
it necessary for properly assessing the validity of our results, for
the following reason: the timescale of variation of the solution is
initially very slow and gradually accelerates as the material falls
into the hole. A convergence analysis for a small total time will not
cover quantitatively the most demanding part of the computation, which
may span a total time of only a couple of decades in $M$, occurring
several hundreds of $M$ {\em after} the computation commences.

\begin{figure}[tbh]
\centerline{\psfig{figure=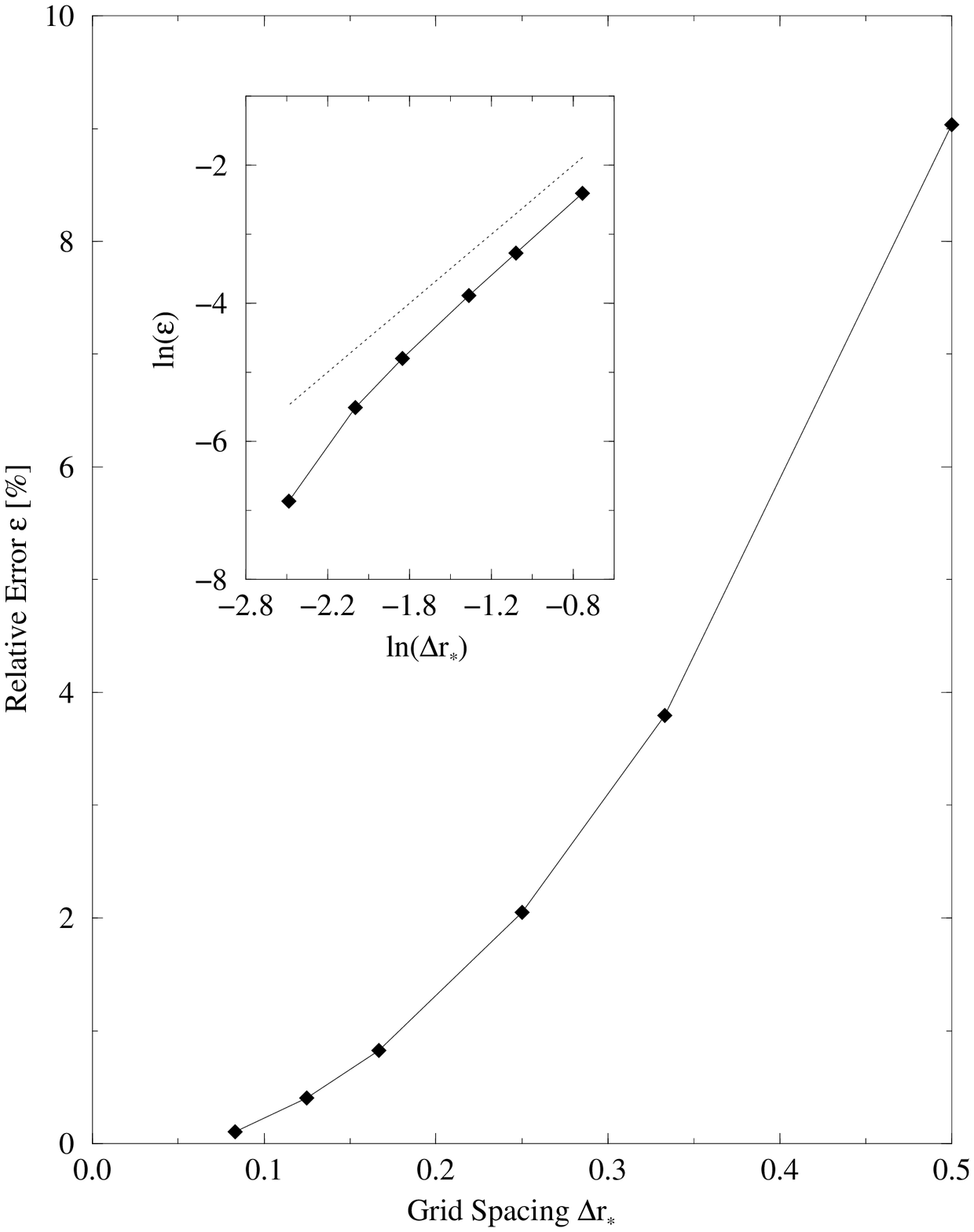,width=6.0in,height=8.0in}}
\caption[Figure 1.]{
\label{conv-vacuum} 
Calibration of the numerical error in computing the {\em scattering
and propagation} of gravitational waves. Plotted is the percent
relative error in the total emitted energy associated with a pure
disturbance (i.e., no matter sources), after $400M$ of evolution. The
insert illustrates the Cauchy converge to a reference grid of $800
\times 20$ points. The dotted line shows for comparison the
theoretical second order convergence rate. The measured rate is about
2.2.  The wave propagation converges relatively fast: fewer than 400
radial points within $50M$ from the black hole, are sufficient for a
relative error of less than one percent.}
\end{figure}

Figure~\ref{conv-vacuum} calibrates the resolution requirements of the
wave code in {\em propagating} radiation on a black hole
background. What is shown is the percent relative error $\epsilon$ in
the total outgoing flux as a function of grid size, i.e., $\epsilon =
(E_{n} - E_{\infty})/E_{\infty}$, where $E_{n}$ is the emitted energy
for a given radial grid size $n$ and $E_{\infty}$ is the energy
out-flux estimated using the finest grid. The flux is integrated at
$r_{*}=100M$ up to a final time $t=400M$. The ``vacuum'' initial data,
for this set of runs were chosen as
\begin{eqnarray}
Z       & = &  e^{-0.2 (r_{*}-25)^2}  \sin^2\theta \,,\\
\dot{Z} & = &  0 \,.
\end{eqnarray}
The resolution requirements are modest, the slope of convergence of
the total energy is at about $2.2$. The convergence curve
overestimates the rate when the grid size becomes comparable to the
reference grid, so for the computation of a numerical convergence rate
we dropped the second finest resolution.

\begin{figure}[tbh]
\centerline{\psfig{figure=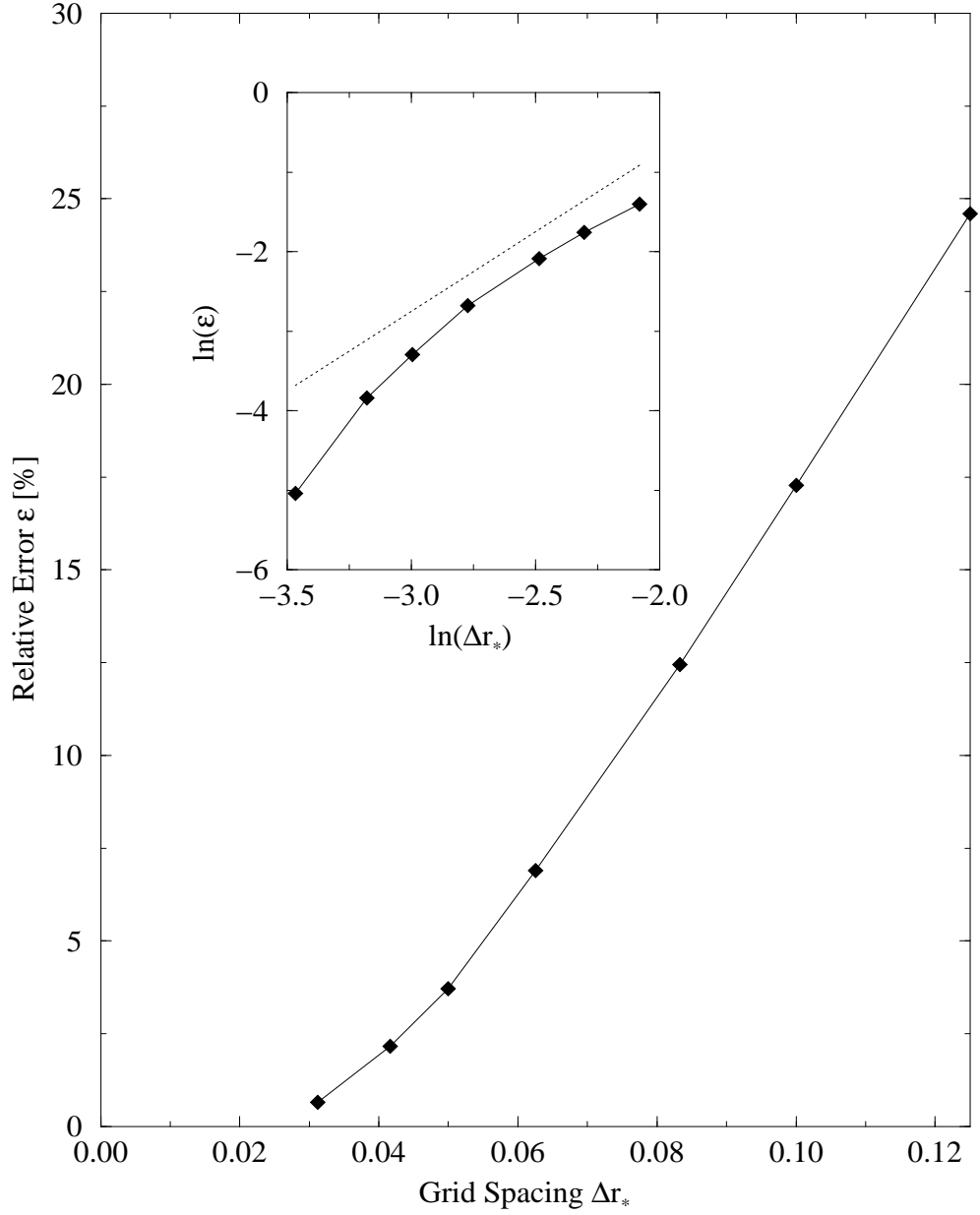,width=6.0in,height=8.0in}}
\caption[Figure 2.]{
\label{conv-shell} 
Calibration of the numerical error in computing wave {\em generation}
by accreting matter. Plotted is the percent relative error in the
total emitted energy associated with an infalling shell (see text for
precise parameters) after $400M$ of evolution. The insert again
focuses on the rate of Cauchy convergence of the total energy,
computed with a reference grid of $2000 \times 20$ points. The rate is
measured to be about 1.9.  The resolution requirements for achieving
levels of error comparable to those of Fig.~\ref{conv-vacuum} are now
considerable larger: 1200 zones inside $50M$ are needed to achieve an
energy estimate accurate to about 2.5 percent.}
\end{figure}

In Fig.~\ref{conv-shell}, the considerably more demanding convergence
behavior for a complete accretion/emission event is demonstrated.  We
consider a typical Gaussian quadrupolar shell, fixing the relevant
parameters to an initially stationary configuration located at $r_{*}
= 35M$, and width $k = 2$. Details about the initial data and the
waveform generated by the infalling matter are given in
Sec.~\ref{results}. As with the free wave propagation, we measure the
energy flux at $r_{*}=100M$ up to a final time $t=400M$ and compute
the total energy. The wave initial data are zero for both the field
and its derivative. The radial extend of the hydrodynamical grid is
between $-3<r_{*}<50$. The wave grid extends between $-50<r_{*}<500$.

Resolutions of $0.2M$, while more than adequate for capturing typical
propagating waves and their interaction with the black hole potential,
seem to be rather inaccurate in resolving the peak emission event. The
plot suggests that in terms of adaptive grids (with a refinement ratio
of two), (e.g.,~\cite{amr-1}) about two levels of refinement around
the hole would substantially decrease the computational demands of
three-dimensional calculations (assuming that a uni-grid resolution of
$0.2M$ can be achieved far from the hole).

%
\section{Results of axisymmetric radial infall simulations}
\label{results}
%

\subsection{Radially Imploding Shells: Isotropic velocities}

\subsubsection{A typical evolution}

The different initial data sets we explore are parametrized by: i) the
location of the shell, $r_0$, ii) its peak density, $\rho_{max}$, iii)
its peak velocity $v_{max}$ and iv) its width parameter, $k$,
\begin{eqnarray}
\rho & = & \rho_0 + \rho_{max} e^{-k(r_{*}-r_0)^2} g(\theta) \,,\\
v    & = & - v_{max} f(\theta) \,,
\end{eqnarray}
where $\rho_0$ is a given background density, of very small value
(typically three to four orders of magnitude smaller than the peak
density). The angular structure encoded into the functions
$g(\theta)$, $f(\theta)$ is modelled using simple spherical harmonics
of low multipole number, e.g., $l=0,1,2$. As mentioned before, we
restrict our discussion to dust ($p=0$). The sign of the velocity
reflects our further restriction to inflows.

\begin{figure}[tbh]
\centerline{\psfig{figure=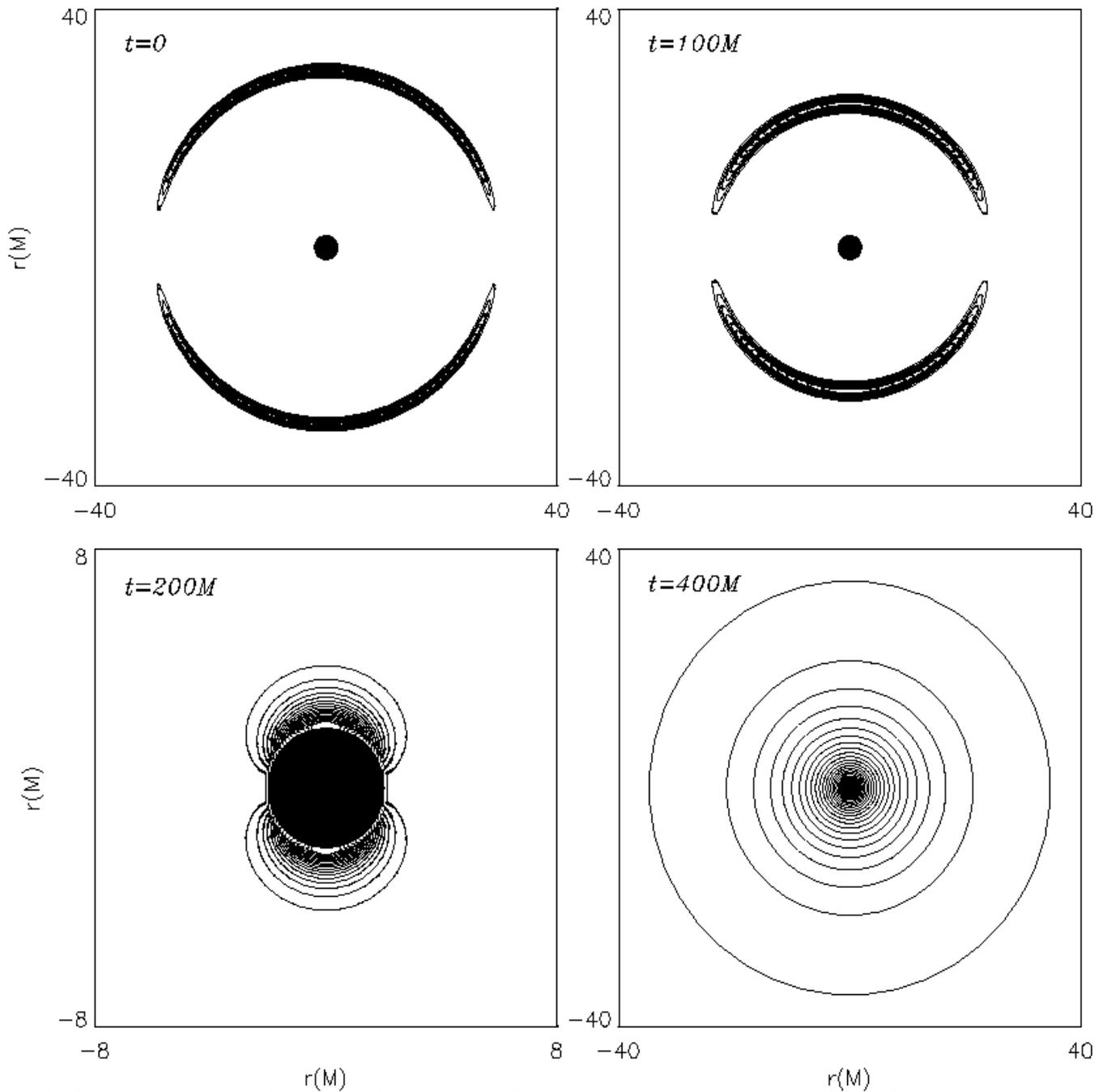,width=7.0in,height=7.0in}}
\caption[Figure 3.]{
\label{quad-dens} 
Snapshots of density for a typical shell accretion computation. The
symmetry axis $\theta=0$ is horizontal, and the location of the black
hole is depicted with the small black disk at the center. The initial,
radially very compact profile of the density is shown at $t=0$,
followed by the gradual acceleration and infall shown at times
$t=100M$ and $t=200M$. The final state is a relic (non-radiative)
spherical pattern, corresponding to the low background density that
``fills the vacuum'' surrounding the shell. The $t=200M$ panel is a
four-fold zoom into the interesting region. The lines are iso-density
contours}
\end{figure}

We present first an overview of a typical evolution. The initial
distribution of matter density is described by a shell with
quadrupolar angular shape, i.e., $g(\theta)=\sin^{2}\theta$ (see the
upper left panel in Fig.~\ref{quad-dens}). The initial velocity
$v_{max}$ is assumed zero. Under the action of the black hole
potential, the flow will develop into an isotropic radial velocity
field. The quadrupolar angular dependence for $\rho$ is the simplest
one that would lead to emission of radiation under isotropic radial
infall conditions.  The shell is initially centered at
$r_{0}=r_{*}=35M$. The Gaussian parameter is $k=2$.

\begin{figure}[tbh]
\centerline{\psfig{figure=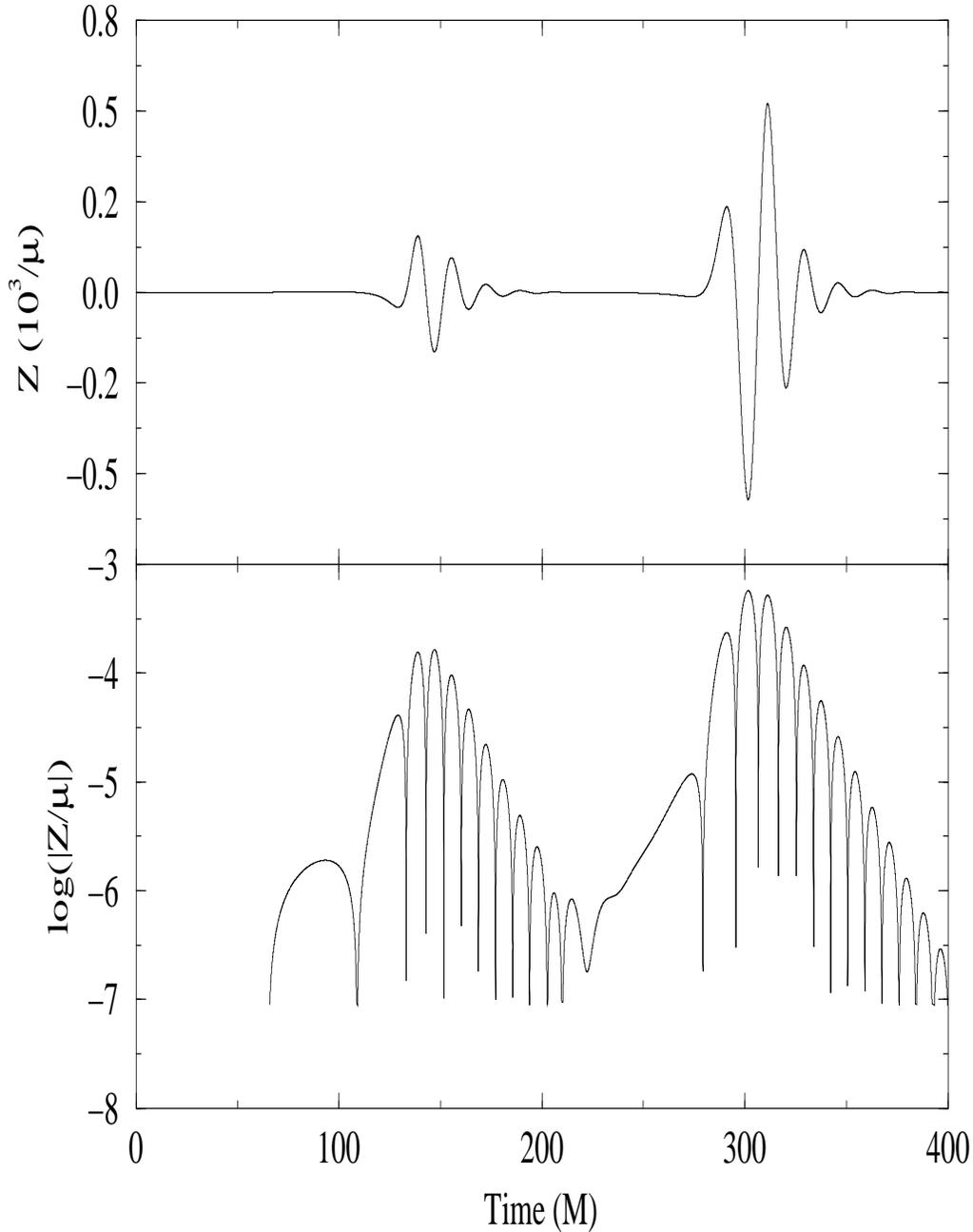,width=6.0in,height=8.0in}}
\caption[Figure 4.]{
\label{time-signal} 
The gravitational wave time signal of a typical shell accretion
event. The upper panel shows the quadrupole part of the wave signal
normalized by the total mass of the shell, reaching a remote observer
($r_{*}=100M$). The presence of two bursts is the first striking
observation. The second burst is more than two times stronger, but the
wavelength appears to be comparable. To bring out the full range of
features, we show in the lower panel the logarithm of the same
signal. The nature of the signal is now clearer: besides initial low
frequency features ahead of each burst, both bursts are mostly in the
quadrupole ringing frequency of the black hole ($17M$). Notice the
transition region in between the two bursts; a non-oscillatory
precursor wave is seen to rise steadily in amplitude, finally giving
way to the strong oscillatory emission}
\end{figure}

The upper panel of Fig.~\ref{time-signal} shows the wave signal, more
precisely the quadrupole part of $Z$ normalized with the total mass of
the shell, reaching a remote observer (located at $r_{*}=100M$) as a
function of coordinate time. The existence of two bursts is the first
striking observation. The second burst is more than two times
stronger, but the wavelength appears to be comparable. To bring out
the full range of features, we show in the lower panel the logarithm
of the same signal. The nature of the signal is now clearer: besides
initial low frequency features ahead of each burst, both bursts are
mostly in the quadrupole ringing frequency of the black hole ($17M$).

The existence of two bursts asks for some clarification. To start, we
notice that $140M$ is approximately the time it takes for initial
radiation to propagate from $r_{*}=35M$ to about $r_{*}=5M$, scatter
off the combined angular-momentum black hole potential, and reach back
to the observer at $r_{*}=100M$. The second burst occurs much latter.
Starting from rest at $r_{*}=35M$, the free fall time of a particle
would be around $200M$, which combined with the $100M$ to reach the
observer, gives us the arrival time of the interesting emission,
associated with the shell crossing the peak of the gravitational
potential and subsequently accreting onto the hole.

\begin{figure}[tbh]
\centerline{\psfig{figure=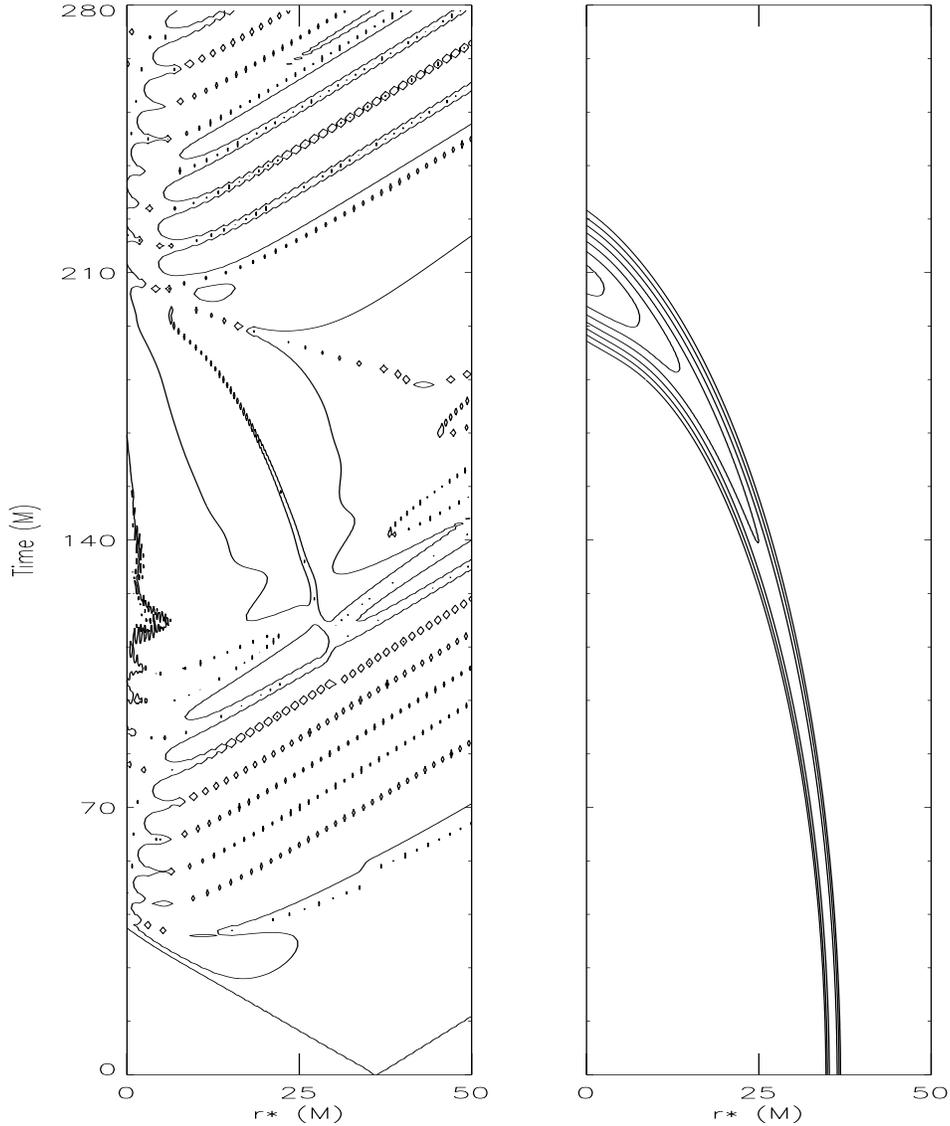,width=6.0in,height=7.0in}}
\caption[Figure 5.]{
\label{space-time} 
Spacetime diagram of wave and density evolutions. Light rays in the
diagram are propagating at $45^{o}$ angles. The domain depicted
extends only to $r_{*}=50M$, in order to bring out the interesting
inner region features. The field variables are sampled at the equator
($\theta=\pi/2$). Both panels show the contour lines of equal
amplitude, for the wave signal (the logarithm of $Z$) on the left, and
the logarithm of density on the right. In the right panel we see quite
clearly the parabolic infall trajectory of the shell. The left panel
shows the prodigious response of the black hole curvature to the
presence of the shell. There are only three contour levels illustrated
in the left panel, at logarithmic amplitudes of $-6,0,2.5$
respectively. The, lowest, $-6$ level is seen to form a light cone
emanating from the initial location of the shell. The cone of
influence reaches inwards to the horizon region, and excites
ringing. The features of this ringing are traced summarily by the $0$
contour level, e.g., the ringing is fully developed at around time
$t=70M$.  As this initial burst subsides exponentially, the shape of
the infalling shell is emerging, through the {\em imprint} it leaves
on the curvature, as can be seen in the interval from $120M$ up until
$210M$. At $t=210M$ the shell accretes rapidly, and it quasi-static
curvature pattern gives way to strong emission (traced by the $2.5$
contour), which is seen to be here predominantly in QNM form.}
\end{figure}

A spacetime diagram (Fig.~\ref{space-time}) of the whole evolutionary
sequence makes the previous discussion fairly transparent. The use of
the tortoise coordinate implies that light rays in the diagram are
(modulo graphic distortions) propagating at $45^{o}$ angles.  Some
guidance to the plot may be worthwhile: the radial coordinate is
truncated at the fairly small value $r_{*}=50M$ to bring out the
interesting features. The field variables are sampled at the equator
($\theta=\pi/2$). We show {\em contours} of equal amplitude. In the
left panel the quantity contoured is the logarithm of $Z$, whereas the
right panel depicts the logarithm of the density. The right panel
illustrates the infall of the shell, imperceptibly at first, then
increasingly faster. Notice the wide spread (in coordinates!) of the
accretion event. The left panel shows the response of the black hole
curvature to the trajectory of the shell. There are three contour
levels illustrated, at logarithmic amplitudes of $-6,0,2.5$
respectively. The $-6$ level is seen to form a light cone emanating
from the initial location of the shell. We can clearly see the cone of
influence reaching inwards to the horizon region, and exciting
quasi-normal mode (QNM) ringing. The features of the ringing are
traced, e.g., around time $t=70M$, by the $0$ contour level. As this
initial burst subsides exponentially, the shape of the infalling shell
is emerging, through the imprint it leaves on the curvature (still $0$
level contours, starting at around $t=120M$). At $t=210M$ the shell is
rapidly accreting, its quasi-static curvature pattern giving way to
emission, which is again seen to be predominantly in QNM form.

The important fact we deduce from this figure is that the emission we
observe is independent of the choice of initial wave data provided the
collision occurs after a total time $t_{c}+t_{r}$, where $t_{c}$ is
the light crossing time of the combined object and $t_{r}$ is the time
interval in which the black hole ringing decays significantly below
the level of the subsequent emission. With a conservative assumption
of similar amplitudes for unphysical and physical emission, the
exponential decay of the QNM suggests (see e.g., lower panel of
Fig.~\ref{time-signal}), that $t_{r}=50M$ should give two orders of
magnitude leeway.

In Fig.~\ref{quad-dens} we plot the spatial density distribution at
four different times. The symmetry axis $\theta=0$ is horizontal, and
the location of the black hole is depicted with the small black disk
at the center. For the plot we use a simple mapping of a meridian
plane of the \sch black hole into ``Cartesian'' coordinates:
$\tilde{x}=r\cos\theta$, $\tilde{y}=r\sin\theta$. The initial,
radially very compact, profile of the density is shown at $t=0$,
followed by the gradual acceleration and infall shown at times
$t=100M$ and $t=200M$. The final state is a relic (non-radiative)
spherical pattern, corresponding to the low background density that
``fills the vacuum'' surrounding the shell. The $t=200M$ panel is a
four-fold zoom into the interesting region.

\begin{figure}[tbh]
\centerline{\psfig{figure=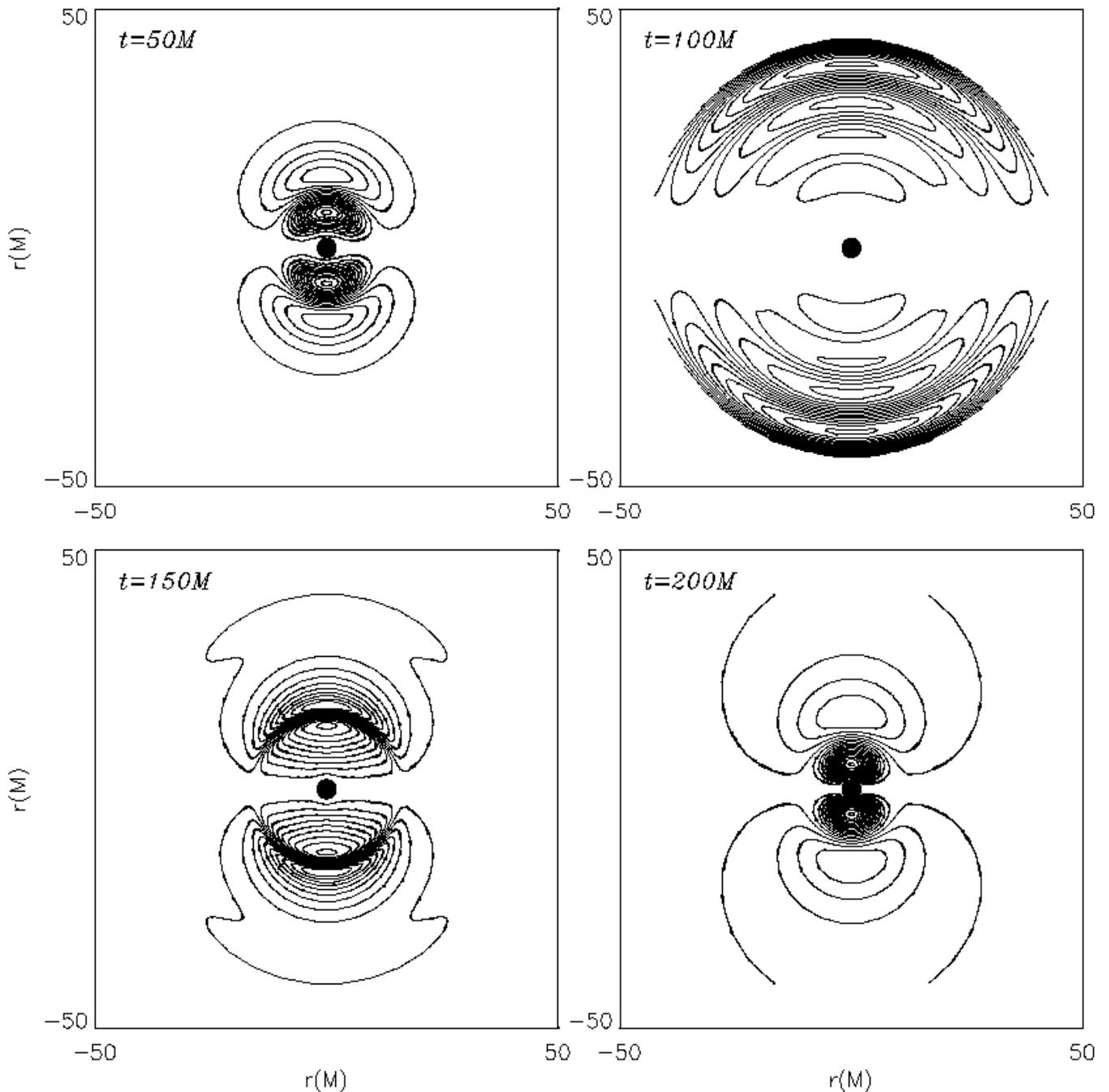,width=7.0in,height=7.0in}}
\caption[Figure 6.]{
\label{quad-wave1} 
Spatial snapshots of wave profiles for a typical accretion event (part
1). The panels can be thought of as horizontal slices of the left
column in Fig.~\ref{space-time}, restoring one suppressed dimension at
the corresponding time levels. The initial burst and associated
ringing are shown at $t=50M$ and $t=100M$. At $t=150M$ those initial
features have subsided sufficiently to reveal the quasi-static field
surrounding the shell and being dragged with it towards the hole. At
$t=200M$, the density is near the peak of the potential (compare to
Fig.~\ref{quad-dens}), and the emission is reaching its climax. }
\end{figure}

\begin{figure}[tbh]
\centerline{\psfig{figure=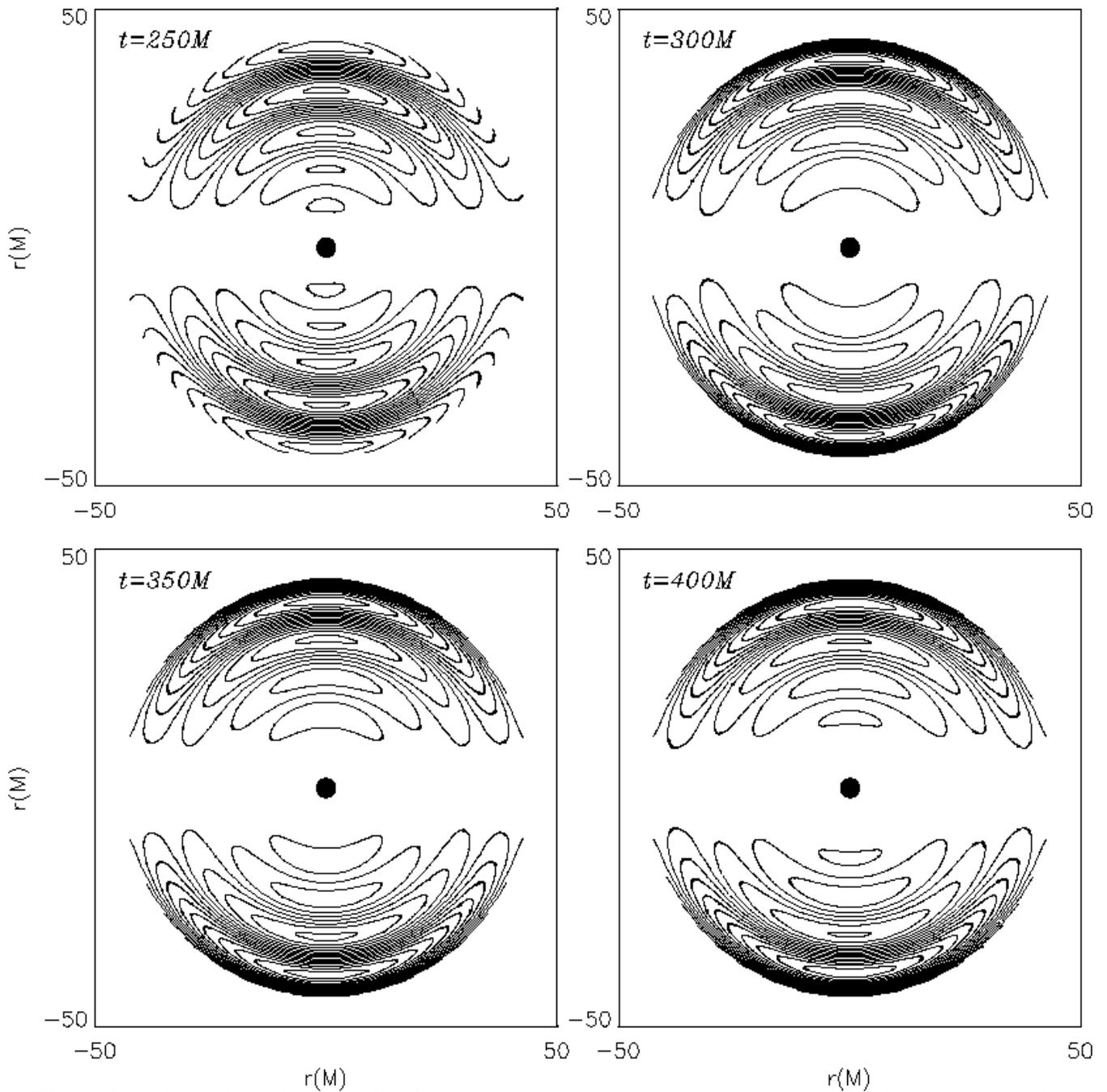,width=7.0in,height=7.0in}}
\caption[Figure 7.]{
\label{quad-wave2} 
Spatial snapshots of wave profiles for a typical accretion event (part
2).  At $t=250M$ the excitement is already gone: the remaining
evolution through $t=400M$ only obtains the slow decay of the black
hole quasi-normal mode ringing. The black hole returns to its
quiescent state.}
\end{figure}

Similarly, we plot spatial patterns for the response of the curvature
in Figures~\ref{quad-wave1} and~\ref{quad-wave2}. The initial burst
and associated ringing are shown at $t=50M$ and $t=100M$. At $t=150$
those features have subsided sufficiently to reveal the quasi-static
field surrounding the shell and being dragged with it towards the
hole. At $t=200M$, the density is near the peak of the potential and
the emission is reaching its climax. At $t=250M$ the excitement is
already gone: the remaining integration through $t=400M$, only obtains
the slow decay of the black hole quasi-normal mode ringing.

\subsubsection{Dependence on shell width, velocity and position}

The narrow radial profile shown in Fig.~\ref{quad-dens} was picked to
produce a loud and clear bang. We illustrate here the nature of the
emission as a function of radial width, by varying the parameter $k$
of the density data. The initial physical size of the shell, i.e., its
thickness in terms of proper radial distance is given approximately by
$ L_{*} = 2 \alpha_{c}/\sqrt{k}$, where $\alpha_{c}$ is the lapse at
the center of the shell.  This length defines the location where the
shell density falls to the $1/e$ of its peak value $\rho_{max}$. All
other parameters stay the same.

\begin{figure}[tbh]
\centerline{\psfig{figure=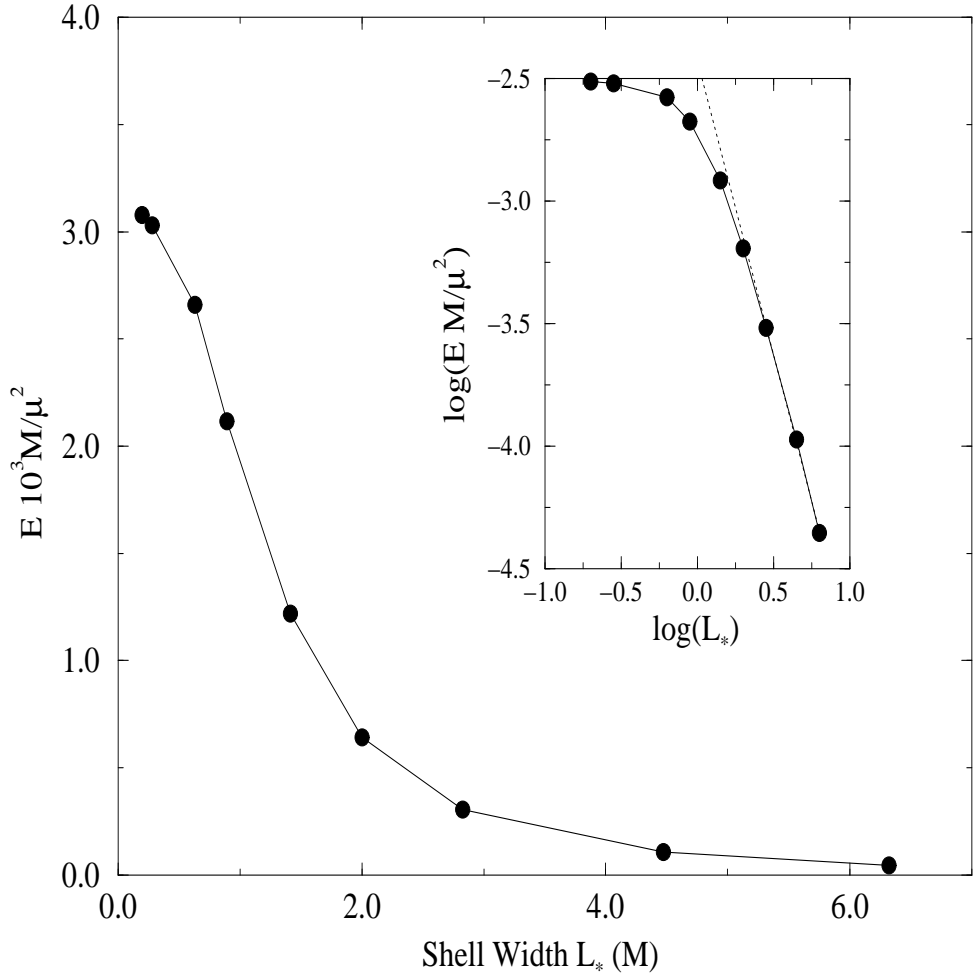,width=6.0in,height=6.0in}}
\caption[Figure 8.]{
\label{s-curve} 
Emitted energy as a function of the radial width of the infalling
shell. More precisely, the energy dependence is shown with respect to
the e-folding (proper) radial length for the shell density. All runs
are for shells centered around $r_{*}=35M$, initially at rest, with
the $k$ parameter ranging from $0.1$ to $100$. The increased
efficiency as a function of compactness is evident, and spans several
orders of magnitude. In the extended limit the energy is shutting off,
apparently as a power law of $L_{*}$ (see insert). The slope of the
power law is fitted well (dashed line) by a value of $-2.4$. In the
thin shell limit, the energy asymptotes to a finite value which is
about a third of the point particle limit. }
\end{figure}

We perform a number of runs, with $k$ parameters ranging from $0.1$ to
$100$ and compute~\cite{energy-comp} the total emitted energy by
integrating in time the outgoing flux at the location of the observer
($r_{*}=100M$).  The dramatic rise in efficiency with increased
compactness is evident in Fig.~\ref{s-curve}, and spans several orders
of magnitude. In the extended shell limit, the energy is shutting off,
apparently as a power law of $L_{*}$. The slope of the dotted line in
the insert is very close to $-2.4$. In the infinitesimally thin limit,
the energy asymptotes to a finite value, about a third of the point
particle upper limit. This is consistent with semi-analytic results
for oblate and prolate spheroids in~\cite{sana81}. From the insert we
deduce that there is a characteristic, initial, radial size, of about
$L_{*}=1$, beyond which no significant efficiency gains are made.

\begin{figure}[tbh]
\centerline{\psfig{figure=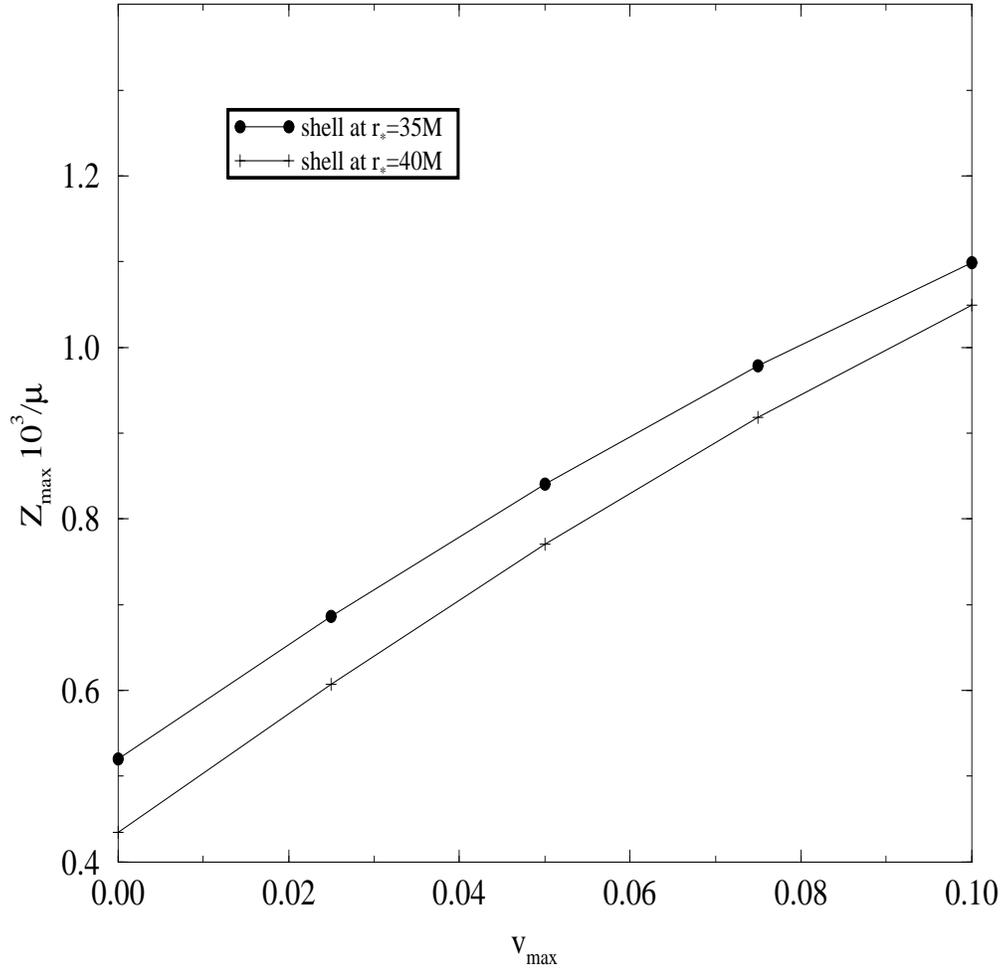,width=6.0in,height=6.0in}}
\caption[Figure 9.]{
\label{velo} 
Emitted energy as a function of the initial velocity of the infalling
shell. Shown is the maximum signal amplitude as a function of initial
velocity, for velocities up to 0.1 times the speed of light. The
infall velocity is seen here to have an important effect on the
emission efficiency,; the amplitude for the largest initial velocity
is more than twice than starting from rest.}
\end{figure}

The velocity of the infalling matter, at the time of crossing the
potential peak, may be expected to be another factor governing the
efficiency of the emission, intuitively anticipating larger impact
velocities to radiate more.  In Fig.~\ref{velo} we highlight this
aspect by plotting the maximum signal amplitude as a function of
initial velocity. This is done for initial velocities up to 0.1 times
the speed of light. The amplitude for the largest initial velocity is
more than twice that for starting at rest. Fitting a law to the
velocity dependence seems problematic without a more delicate
analysis.  The stumbling block is the fact that starting from rest at
a finite distance already corresponds to large velocities at impact,
hence the offset and slope at $v=0$ of our diagram are fairly
arbitrary and do not capture the asymptotic behavior of small
velocity.  The same figure displays a parallel parameter survey for
shells falling from a larger distance ($r_{*}=40M$). The Gaussian
parameter $k$ has been kept the same.  This explains the, at first
confusing, weaker emission from the more remote shell
location. Clearly, compactness is overpowering velocity (at least for
the range of configurations probed here) in controlling the
efficiency.

\subsubsection{Excitations of QNM emission}

\begin{figure}[tbh]
\centerline{\psfig{figure=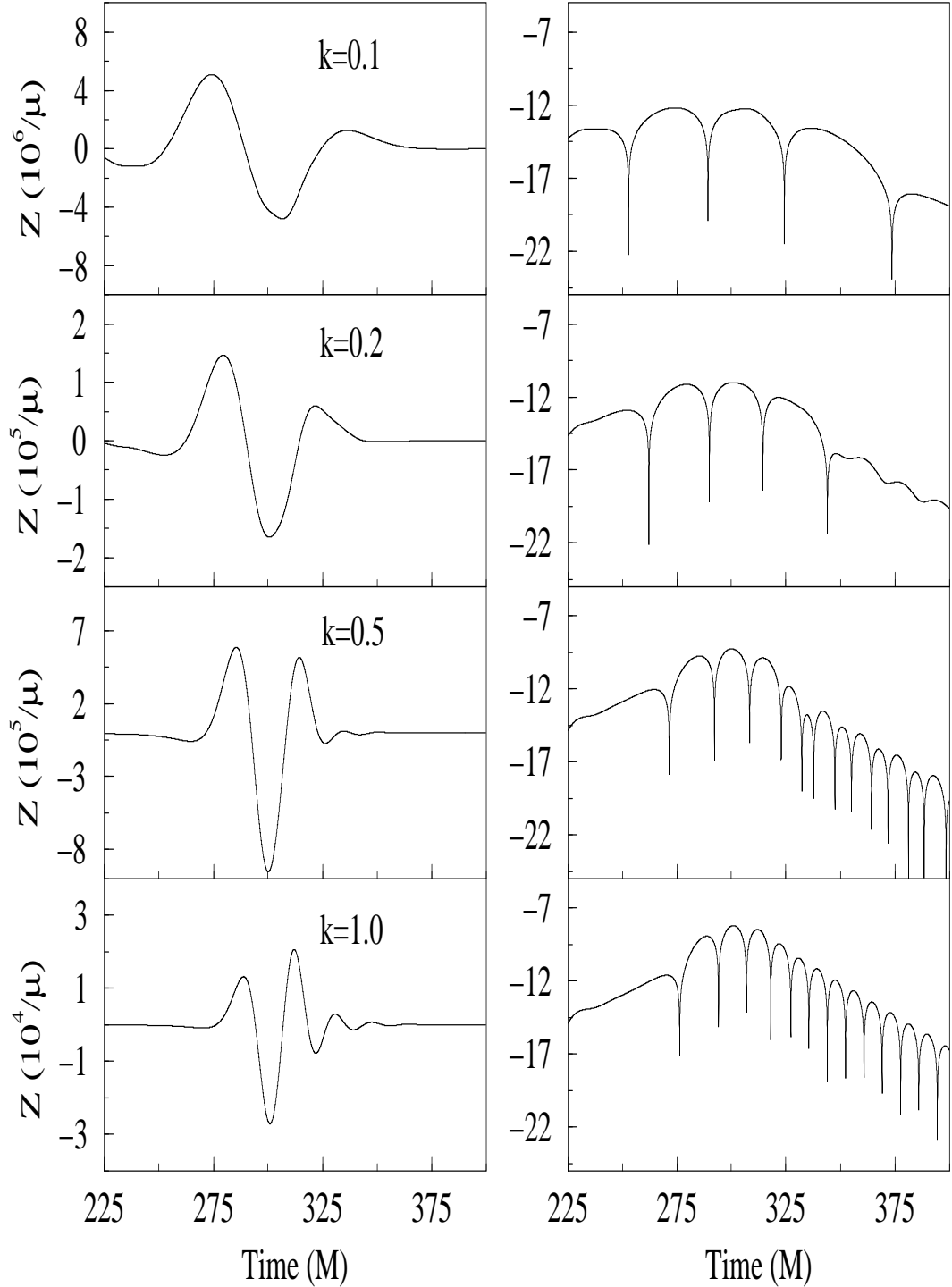,width=6.0in,height=8.0in}}
\caption[Figure 10.]{
\label{excite} 
Excitations of the fundamental (l=2) black hole quasi-normal mode, as
a function of the infalling shell width. The left column depicts the
waveform for four select values of shell width (increasingly thinner
shells from top to bottom).  The large increase of per unit mass
emission is evident. The wavelength of the emission is also
increasingly shorter. The right column shows the logarithm of the
signal, and brings out the qualitative changes emerging in the thin
shell limit. More compact shells are seen to be first marginally
(second panel), and then more clearly, exciting the QNM ringing. In
the last panel the ringing is QNM frequency is dominant from $t=315M$
onwards, but it appears that the maximum amplitude is still emitted at
a somewhat lower frequency.}
\end{figure}

Fig.~\ref{excite} illustrates the {\em qualitatively} different
response of the black hole to the infalling shell, depending on the
shell width. The left column depicts the waveform for four select
values of $k$ (increasingly thinner shells from top to bottom). The
large increase of per unit mass emission is evident. The apparent
wavelength of the emission is also seen to be increasingly
smaller. The right column takes the natural logarithm of the signal,
to bring out the qualitative changes brought about by the changing
shell profile. With progressively more compact shells we see first a
marginal (second panel), and then a more clearcut (third panel),
excitation of the QNM ringing~\cite{ringing}. In the last panel the
ringing is dominant from $t=315M$ onwards. Hence, as intuitively
anticipated, a broad initial distribution, emits less, in longer
wavelengths, and correspondingly does not excite the lowest QNM
frequency of the black hole as much.

The importance of QNM waveforms in the interactions of black hole with
external matter was illustrated early on~\cite{cunningham}.  More
detailed numerical work~\cite{eseidel} corroborated the case for QNM
excitation in stellar collapse. On the other hand, calculations based
on point particles {\em scattering} of a black
hole~\cite{oona84a,kojima} showed excitations induced by an external
object that do not produce significant QNM ringing. In subsequent
work~\cite{sun-price} the amount of excitation of the QNM modes during
core collapse has been examined using both analytic model problems and
further simplified numerical calculations in the spirit
of~\cite{cunningham}.

Despite the considerably different setup, it appears that our findings
support the qualitative and intuitively agreeable statement, made
in~\cite{sun-price}, that QNM excitation is induced by the curvature
profiles that have {\em spatial wavelengths} comparable to the width
of the black hole potential. In our setup, the width of the shell
controls the shape of the quasi-static curvature profile. Modulating
the profile using the width parameter, strongly affects the degree of
excitation of the QNM signal.

\subsection{Radially Infalling Shells: Anisotropic Velocities}

One of the physical objectives in this paper is to approximately map
the role of the fluid source term in the case of radial infall. It
seems that anisotropic velocity profiles have not been investigated in
the literature to the extent that isotropic flows have. From the
structure of the source term we see that in this case the {\em source
non-linearity} of the right-hand-side of the BP equation becomes
important. Quadrupolar (and of course higher) contributions may emerge
from couplings of angular structure between velocity and density
profiles.

\subsubsection{Spherical shell collapsing onto a moving black hole}

The geometrically simplest, {\em dipole}, anisotropy in the velocity
field occurs naturally for matter motion with {\em net linear
momentum} with respect to the black hole. One interesting scenario is
the accretion of the layers of a star onto a core that has acquired
net momentum (kick velocity) during the collapse. To simulate such a
configuration we consider a spherical Gaussian distribution of density
centered at $r_{*}=35M$ (simulating a dense layer of material, with
its various parts initially at rest with respect to each other). We
assume that earlier events have given the core of the star (now
collapsed into a black hole), a linear velocity of about $0.1c$. In
practice, the computation is done in the rest frame of the black hole,
and the linear momentum of the fluid translates to a radial velocity
profile $f(\theta)=\cos\theta$. The value we adopt for the linear
velocity is generous compared to the typical observed kick velocities
of {\em neutron stars}, which reach, at best, one-thirtieth of this
speed. The unphysically large value is adopted here for illustration
purposes.

\begin{figure}[tbh]
\centerline{\psfig{figure=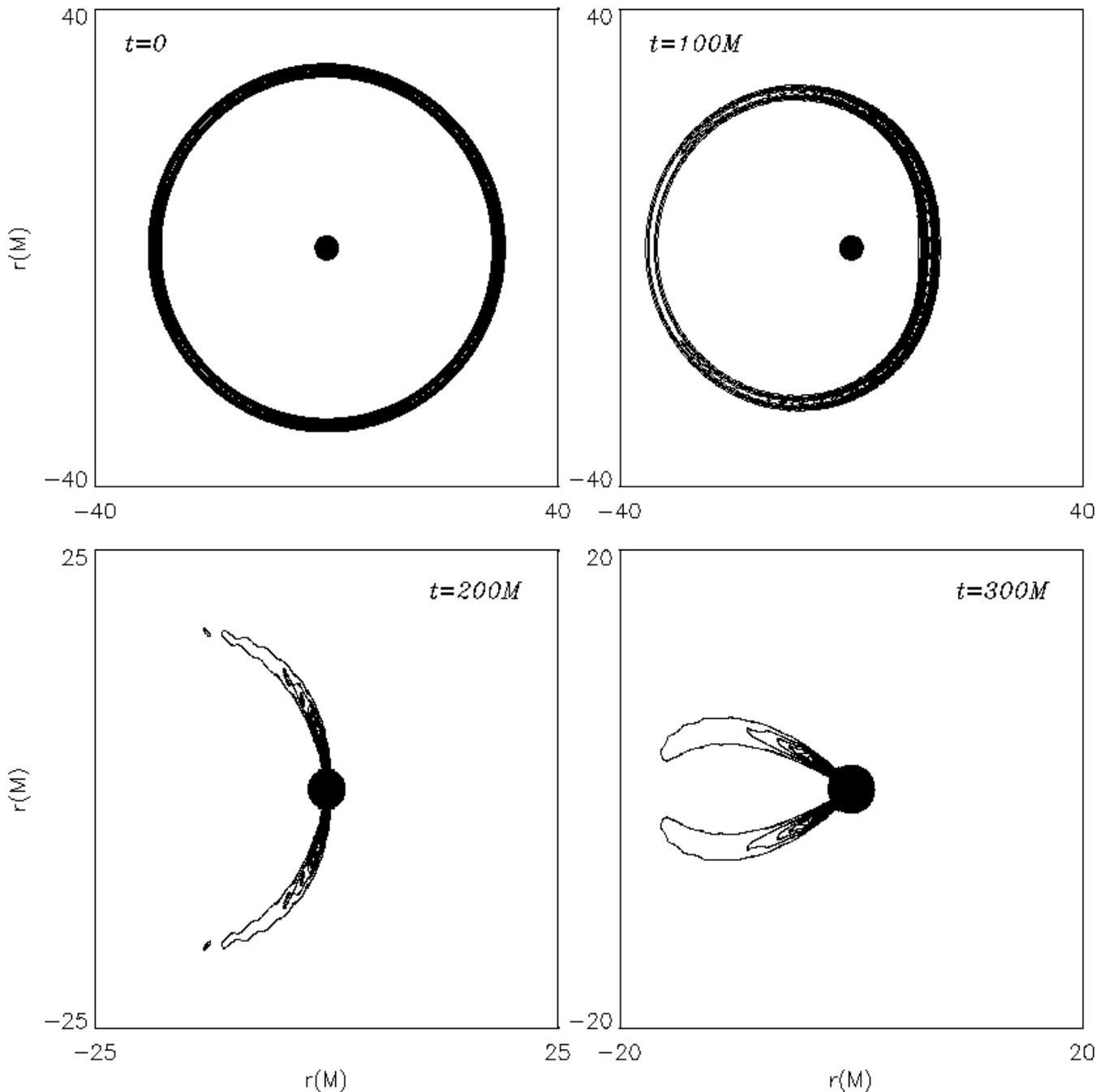,width=7.0in,height=7.0in}}
\caption[Figure 11.]{
\label{kick} 
The density evolution for a shell collapse onto a moving black hole.
The hole rushes up the symmetry axis (right side of the panels). By
$t=100M$ the initially spherical shell has been visibly distorted, as
parts of it experience increased acceleration due to the proximity to
the approaching hole. By $t=200M$, the hole has gone right through the
shell and we can see concentrated accretion to occur at the equatorial
plane. At this time most of the gravitational wave emission has
occurred. In the final snapshot (at $t=300M$) we can see the slow
accretion of the remaining shell material from the back side of the
black hole. The lower panes progressively focus on the inner region.
Despite the already astrophysically unlikely kick velocity of $0.1c$,
the energy released in this event (as $l=2$ radiation) is measured to
be only $2.9 \times 10^{-4}\mu^2/M$. The waveform is very similar to
the bottom panel in Fig.~\ref{excite}}
\end{figure}

Fig.~\ref{kick} shows the development of the density at intervals of
$100M$. The hole rushes up the symmetry axis, towards the positive
hemisphere (right side of the panels). By $t=100M$ the shell has been
distorted appreciably, as parts of it experience increased
acceleration due to the proximity to the approaching hole. By
$t=200M$, the hole has gone right through the shell, and we can see
concentrated accretion to occur at the equatorial plane. At this time
most of the gravitational wave emission has occurred. In the final
snapshot (at $t=300M$) we can see the slow accretion of the remaining
shell material from the back side of the black hole.

The time profile of the quadrupole signal component is essentially
identical with the high compactness shells we considered before (see
e.g., last panel of Fig.~\ref{excite}).  The energy released in this
event as $l=2$ radiation is measured at $2.9 \times 10^{-4}
\mu^2/M$. The numerical error, estimated from the radial calibration
plot (Fig.~\ref{conv-shell}), indicates that the numerical energy
estimate is accurate to about 2\%.

\subsubsection{Pancake collapse}

In the case of rotational collapse one would expect more involved
anisotropic velocity profiles. We do not simulate in this work
rotational configurations, but will attempt to mimic a differential
fall using an anisotropic velocity pattern.  We hence assume again a
spherical Gaussian distribution of density centered at $r_{*}=35M$ and
take $f(\theta)=\sin\theta$.

\begin{figure}[tbh]
\centerline{\psfig{figure=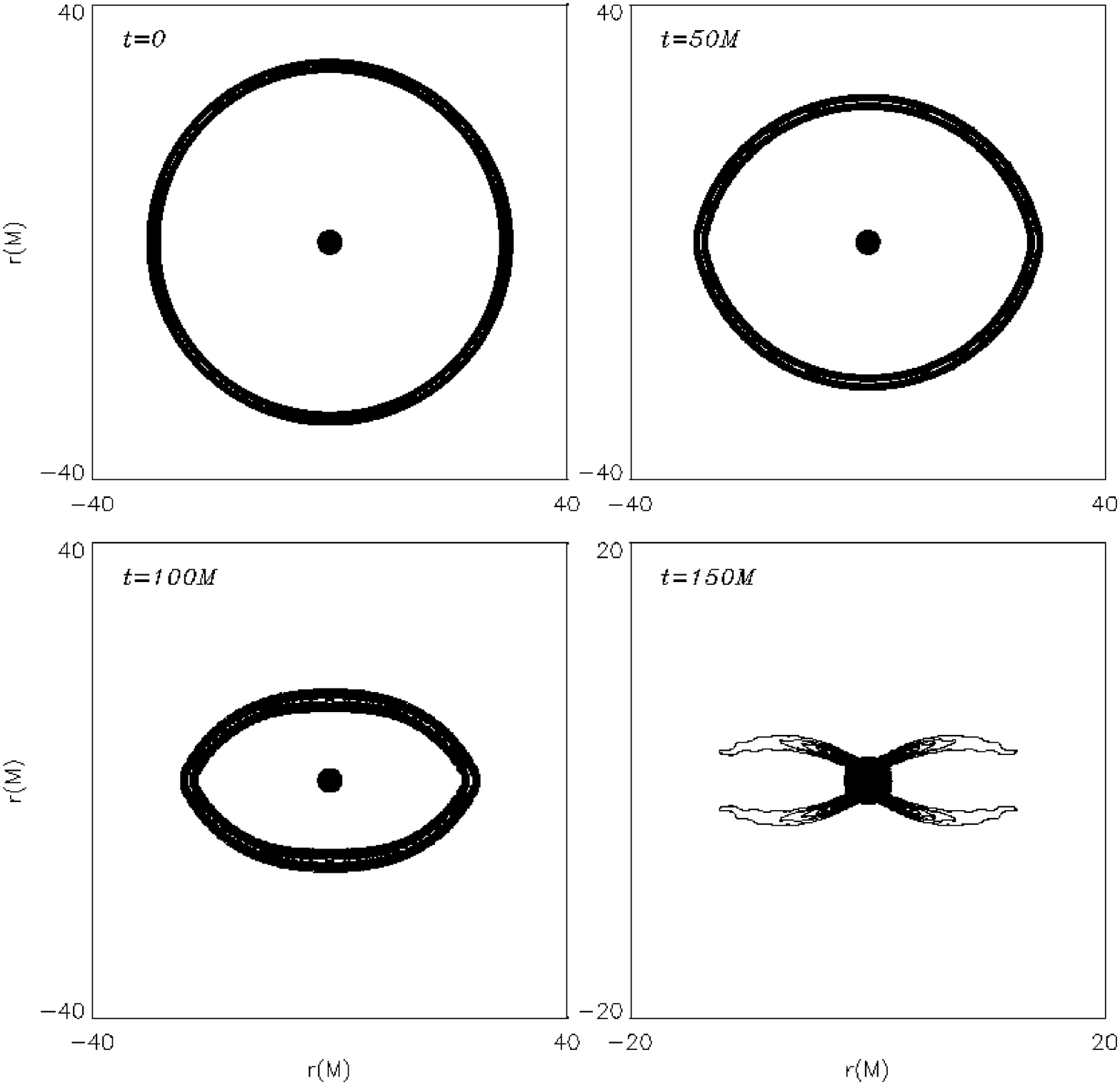,width=7.0in,height=7.0in}}
\caption[Figure 12.]{
\label{pancake} 
The density evolution for anisotropic, ``pancake'', collapse.  The
shell collapses onto the black hole with the equatorial region falling
in first, while the polar regions accelerate much later. By $t=150M$
the equatorial zones have accreted; further material keeps falling in
from larger latitudes. The energy released here in the quadrupole
mode is measured at $2.4 \times 10^{-3}\mu^2/M$, and comes
predominantly from the equatorial accretion event.}
\end{figure}

Fig.~\ref{pancake} shows the development of the density at
intervals of $50M$. The shell collapses onto the black hole 
with the equatorial region falling in first, while the polar
regions only gradually accelerate. By $t=150M$ the equatorial
zones have accreted and further material keeps falling in
from larger latitudes. The energy released here (again in the
quadrupole mode) is measured at  $2.4 \times 10^{-3} \mu^2/M$. 
The accuracy is again at the 2\% level. 

Comparing the efficiency figures for the two anisotropic scenarios
considered here we notice the order of magnitude difference. Given
that the shell material has for both cases similar {\em peak} radial
velocities on impact, it is likely that the key factor in determining
the efficiency is what {\em fraction} of the total mass accretes with
velocities close to the peak and radial profile sufficiently compact.

\section{Discussion}
\label{discuss}

We developed a new method that lies in the borderline between fully
non-linear numerical integrations of the Einstein equations coupled to
matter fields, and the test particle semi-analytical investigations.
Our primary purpose is estimating the gravitational radiation emitted
by complex motions of matter sources in the vicinity of black
holes. We opted to approximate the first order deviations from the
exact black hole geometry using curvature perturbations induced by
perfect fluids whose non-linear evolution is integrated using a
hydrodynamical code. We used typical infall computations to
demonstrate the second order convergence of the signals and calibrate
the energy estimates over a very large amount of evolution time.

We studied dust shells falling isotropically from finite distances,
with variable shell shapes and initial velocities. We explored the
dependence of the energy emission on the shell thickness and initial
velocity, fitting the numerical data when possible. We demonstrated
the gradual excitation of the black hole quasi-normal mode frequency
by sufficiently compact shells, confirming ideas about the mechanisms
responsible for producing strong QNM signals. We extend the existing
literature on radial infall of dust shells by exploring the case of
anisotropic radial infall and the dependence of the signal on the
degree of anisotropy. The anisotropic scenarios we investigate include
the collapse of an initially spherical layer of material onto a moving
black hole and the anisotropic ``pancake'' collapse of a spherical
layer.  In broad terms, in the case of radial infall onto a
non-rotating black hole, the dominant factor controlling the strength
of the emission is, quite generically, the radial compactness of the
source.

We introduced and explored a prescription for minimizing the initial
radiation content which appears to perform rather well in the current
context. The prescription concerns the selection of the {freely
specifiable data}, on the initial hypersurface, and essentially
consists of solving the {\em static problem for the electric and
magnetic parts of the curvature tensor}, derived from their
corresponding evolution equations. In the presence of non-zero matter
sources, the procedure generates the static electric and magnetic
curvatures associated with density distributions and fluid
currents. The approach can be transfered, at least in principle, to
systems of non-linear evolution equations that incorporate some, or
all, components of the Weyl tensor in the evolution variables.

Despite the reduction in the amplitude of the initial burst, the very
low efficiency of the sources we simulated in the present work called
for more drastic action. The remnant radiation in the initial data has
been dealt here in an unambiguous, brute force, way: the computation
simply marches along for sufficient time, past any shadows cast by the
initial data prescription. This illuminates an important facet of
numerical relativity computations: In the absence of convincing
arguments demonstrating minimal contamination from the initial data,
trustworthy signals can only be obtained by evolving for sufficiently
long times. This allows the first burst of radiation to decay and
reach negligible levels which do not obscure the real signal. If this
strategy is adopted, a reduction of the amount of radiation in the
initial data has positive practical consequences, as it reduces the
computational resources required for unambiguous identification of the
signal, especially in three-dimensional (3D) simulations. It is
reassuring though, that the problem appears to become less relevant in
precisely such 3D long term integrations of {\em strong emitters},
such as particles orbiting around black holes~\cite{amr-2}. In that
case, the strong quadrupolar emission of the binary appears to be
quickly dominating the initial burst.

The flow of matter in the vicinity of the black hole, for a major
initial part of the computation, involves timescales at least one
order of magnitude slower than the timescale of propagating
gravitational waves. This circumstance may be encountered quite
frequently when attempting to simulate realistic dynamical events, as
the progenitor (slow) system must be modelled for a sufficient amount
of time to ensure proper initial configuration. This disparity in
timescales implies that explicit methods of integration of the
geometry (as the one employed here) are constraining the efficiency of
the computation. Implicit integrations of the wave degrees of freedom
may offer a more economical alternative, by allowing an integration at
the hydrodynamical timescale. During the initial, slow motion stage,
existing gravitational waves do not need to be resolved accurately in
time, hence the timestep can be increased to match the hydrodynamic
CFL condition.  Monitoring of the fluid velocities, and adaptation of
the time-step during the dynamic phase, will correctly capture the fast
moving phase and the associated waveforms.

As discussed in section~\ref{numerical}, the inner boundary for the
hydrodynamical computation is taken to be further from the horizon
than for the wave equation. In the coordinate system employed here
(Boyer-Lindquist coordinates) the numerical integration of
hydrodynamic equations becomes problematic close to the horizon, as
the Lorentz factor diverges there. We checked that our results are not
affected to a noticeable degree by the location of the boundary, but
it is clearly desirable to have a uniform integration domain for the
two systems, especially in that sensitive and interesting region. A
satisfactory numerical extension of the coordinates {\em inside} the
horizon, for the hydrodynamical equations, has been reported
recently~\cite{HAC_I,HAC_II,HAC_III}. It appears that the computation
of emission from more general flows around, in particular, rotating
holes, will benefit from a complete transcription of the coupled
problem onto regular coordinate patches (and non-singular tetrads).

Radiative boundary conditions at infinity are delicate to impose on a
finite grid, even for the linearized curvature perturbations that we
are considering. Approximate schemes always achieve a partial
transmission of wave energy across the boundary but this transmission
can drop significantly for curved wave-fronts in a three-dimensional
setting. Furthermore, imposing radiative conditions at a finite
distance implies an effective truncation of the equation's
coefficients. This interferes with physical features of the evolution
which depend on terms beyond the principal part, such as quasi-normal
ringing and tails. A developing programme for solving such problems is
the ``Cauchy-characteristic matching''. In \cite{ekgII} this concept
has been applied to equations arising in the context of black hole
perturbation theory. For a review of this entire effort
see~\cite{jeff-living}.  In the present work the problems with
radiative conditions are exacerbated in an unexpected way: whereas for
scalar wave equations incoming and outgoing radiation scales with the
same power of $r$, for the spinorial equations considered here, the
well known peeling properties of the Weyl tensor ensure that this is
not so.  Small amounts of reflection in outer boundaries, propagate
back into the grid and get {\em amplified} as they propagate inwards,
entirely due to the radial scaling properties of incoming waves. In
the absence of a seamless connection to infinity, an expensive, but
safe, way to bypass reflection problems is to consider only the domain
of dependence of the bounded initial hypersurface.

The current implementation of the method allows the study of arbitrary
axisymmetric flows onto a non-rotating hole. Using this setup we plan
to extend the computations performed here to the case of matter
motions with angular momentum and significant pressure support, as for
example the dynamic collapse of the toroidal debris of a disrupted
neutron star onto the black hole companion. Gravitational radiation
emitted in this process would be the swan song of the once powerful
binary emitter. With the parallel development of {\em non-linear}
coupled codes (see e.g.,~\cite{magor}), issues lying at the interface
between perturbation theory and full numerical relativity become
amenable to exploration, most interestingly the effects of self
gravity.

\section{Acknowledgments}
It is a pleasure to thank  
J.M$^{\underline{\mbox{a}}}$.~Ib\'a\~nez, 
P.~Laguna, 
B.~Schmidt, 
and E.~Seidel 
for many helpful comments. P.~P thanks SISSA for warm
hospitality while parts of this work were completed and J.~Miller for
insightful inquiries. J.~A.~F acknowledges financial support from the
TMR program of the European Union (contract number ERBFMBICT971902).
All computations were performed on the SGI Origin 2000 of the AEI.
Finally, we also thank God for our daily brains.


\newpage

\newpage

\newpage

\newpage

\newpage

\newpage

\newpage

\newpage

\newpage

\newpage

\newpage

\newpage

\begin{references}
\bibitem{energy-comp} The integral of the energy emission rises first,
as a result of the initial burst, then levels-off as that emission
decays exponentially, then rises again to register the relevant total
energy. Subtracting the first plateau from the total energy and
normalizing by the total mass gives the efficiency of emission in
units of the black hole mass.
\bibitem{ringing} Our use of the term QNM excitation is loose. It
appears that it is not entirely straightforward to define precisely
how much energy is channelled outwards at the QNM frequency.
\end{references}
\end{document}